\documentclass[a4paper,11pt]{article}
\pdfoutput=1 

\usepackage{jheppub} 
\usepackage{lineno}
\usepackage[T1]{fontenc} 
\usepackage{xcolor}
\usepackage{colortbl}
\usepackage{float}

\usepackage{multirow}
\usepackage{rotating,tabularx}
\newcommand{\myb}{\textcolor{black}}
\newcommand{\myv}{\textcolor{black}}
\newcommand{\myr}{\textcolor{black}}
\newcommand{\myt}{\textcolor{black}}
\newcommand{\myc}{\textcolor{black}}
\newcommand{\myo}{\textcolor{black}}
\newcommand{\myor}{\textcolor{black}}
\newcommand{\mybl}{\textcolor{black}}
\newcommand{\myg}{\textcolor{black}}
\newcommand{\myp}{\textcolor{black}}
\title{\boldmath Maxwell construction and multi-criticality in uncharged generalized quasi-topological black Holes}

\author[a]{Mengqi Lu}
\author[a,b]{Robert B. Mann}

\affiliation[a]{University of Waterloo, \\
200 University Ave W, Waterloo, Canada}
\affiliation[b]{Perimeter Institute for Theoretical Physics, \\ 31 Caroline ST. N., Waterloo, Canada}

\emailAdd{m64lu@uwaterloo.ca}
\emailAdd{rbmann@uwaterloo.ca}
\arxivnumber{2306.06733}

\abstract{we demonstrate the existence of $N$-tuple critical points of uncharged AdS black holes in generalized quasi-topological theories. The criticality is shown to have a geometrical interpretation described by the Maxwell's equal area rule. We present a compact reformulation of the area rule and identify a criterion for the emergence such points. Using this criterion, we construct several multi-critical points with genuine generalized quasi-topological densities, including a quadruple and a quintuple points.
}

\begin{document} 
\maketitle
\flushbottom

\section{Introduction}

 It is well known that black holes behave like thermodynamic systems that radiate a thermal flux of particles \cite{hawking1974black} and whose equilibrium states  are governed by the four laws of black hole mechanics \cite{bardeen1973four}, reinterpreted as the four laws of  thermodynamics.   For 
asymptotically AdS 
 black holes, the negative cosmological constant can be identified as a thermodynamic pressure that validates the extended first law of thermodynamics \cite{creighton1995quasilocal, henneaux1985asymptotically, kastor2010smarr, kastor2011mass}. In this context  black hole mass is interpreted as enthapy, and thus the mechanics of black holes can be understood in   terms of chemical thermodynamics \cite{kubizvnak2015black,kubizvnak2017black}.
 Over the past decade, 
 the perspective of black hole chemistry has led to the discovery of a number of 
 rich properties, including 
 Van
der Waals phase transitions \cite{kubizvnak2012p},
 re-entrant phase transitions \cite{altamirano2013reentrant,frassino2014multiple}, superfluid-like phase transitions \cite{dykaar2017hairy, hennigar2017superfluid, hennigar2017thermodynamics}, and
 triple points \cite{altamirano2014kerr,frassino2014multiple,hull2021thermodynamics,hull2022exotic}.

 A recent interesting discovery in black hole chemistry was that of multi-critical points.  These were first seen in 
 Einstein gravity coupled to non-linear electrodynamics \cite{tavakoli2022multi}, but were shortly afterward found 
 to be present in multiply rotating Kerr-AdS black holes \cite{wu2022multicritical}, and in Lovelock gravity
 \cite{wu2022multicritical2}. In the latter case multi-critical behaviour can even occur for asymptotically flat black holes \cite{wu2022thermodynamically}.
An $N$-th order multi-critical point occurs when $N$ distinct phases merge at a single value of pressure and temperature, generalizing the notion of a triple point (with $N=3$).  
Explicit examples of quadruple and quintuple points have been found in all cases examined so far.

In this paper, we demonstrate an alternative method for finding multi-critical points in black hole phase transtions.   Previous methods exploited the fact that  each extremum of the temperature
(regarded as a function of horizon radius)   corresponded to a cusp in the Gibbs free energy  \cite{tavakoli2022multi, wu2022multicritical2}. 
$2N-2$ distinct extrema were thus required to obtain   $N$ distinct phases, each with its own swallowtail in the Gibbs free energy diagram.
The intersection points of corresponding swallowtails will merge if   two adjacent inflections of $T(r_+)$ occur at the same temperature.  Extending this to all  the inflections will then yield an $N$-tuple critical point where all such intersections merge. However, the number of parameters used in this approach is more than needed -- in fact only approximately half of the horizon radii of the extrema are needed. This redundancy results in extremely low efficiency 
in computing the (finely tuned) parameters need to obtain a multicritical point, as iteratively tuning the various parameters such that all inflections occur at the same temperature  becomes very time-consuming. This  problem  is particularly acute 
when dealing with 
more complicated higher-curvature  theories. Indeed
  the situation deteriorates when considering black holes  with large numbers of thermodynamic  degrees of freedom since considerably high precision is required because of the finely tuned nature of multi-criticality. 
 
The new method  overcomes these difficulties. Inspired by  Maxwell's equal area rule \cite{clerk1875dynamical}, we provide an equivalent but more compact description of multi-criticality, which can be perfectly adapted to the construction without the redundancy introduced in  previous methods.   Instead of iteratively manipulating input parameters, our approach directly indicates whether a given set of non-redundant parameters can or cannot have multi-critical points. If the latter holds, we obtain  their accurate values in thermodynamic  phase space with notably less computation.

To illustrate our method we consider black holes in higher curvature theories of gravity.
Specifically we consider Generalized quasi-topological (GQT) gravity.  The reasons for this are as follows.

For the past 2 decades there has been a revival of interest in higher curvature gravity  in the theoretical physics community. Such theories  have proven to be significant in a variety of contexts in physics, including string theory, holography, the AdS/CFT correspondence, tests of general relativity, and black hole thermodynamics. The gravitational action becomes renormalizable when supplemented with higher-curvature terms  \cite{stelle1977renormalization}, making such theories candidates for a quantum theory of gravity. 
String-theoretic versions of quantum gravity motivate the
possibility of higher dimensional spacetimes, and the addition of higher-curvature corrections allows for broader
generalizations of the Einstein-Hilbert action to
dimensions larger than four. These  higher-curvature theories provide toy models for studying the AdS/CFT correspondence and allow for holographic study of Conformal Field Theories (CFTs). 

But the inclusion of higher order curvature terms comes at a price -- it can yield equations of motion containing higher derivatives in the metric that give rise to instabilities and negative energy modes \cite{woodard2015ostrogradsky, OstrogradskyMmoiresSL}.  Notably, this inconsistency with general relativity (GR) is absent in some classes of higher-curvature theories \cite{lovelock1970divergence,lovelock1971einstein,sotiriou2010f,woodard2007avoiding} in which only a massless spin-2 graviton can propagate to infinity . This subclass of higher-curvature gravity theories is considerably more promising than the others and thus warrants further investigation. 

Generalized quasi-topological (GQT) gravities, a class of recently proposed higher-derivative theories, satisfies the requirements noted above. Theories in the GQT class characterize generalizations of GR in any dimension and to any order in curvature insofar as they contain  non-hairy black hole solutions and  second-order-differential equations for the metric in any linearized maximally symmetric background \cite{bueno2019generalized, Bueno:2022res, Moreno:2023rfl}. Generally, the bulk part of their action can be written as 
\begin{equation}\label{action}
    I_{\mathrm{bulk}}=\frac{1}{16\pi}\int_{\mathcal{V}} \mathrm{d}^dx\sqrt{-g}\bigg[-2\Lambda+R+\sum_{n=2} \sum_k \alpha_{n,k}\mathcal{S}_n^{(k)}\bigg],
\end{equation}
where $\Lambda$ is the cosmological constant, $\mathcal{S}_n^{(k)}$ are independent densities constructed from different constructions of $n$ Riemann tensors and the metric,   $\alpha_{n,k}$ is the associated $k$th higher-curvature coupling, and the Newtonian constant $G$ is set to $1$ for simplicity. Quite remarkably, in the context of gravitational effective field theory, any higher-curvature theory can be mapped into a subset of GQT theories via field redefinition \cite{bueno2019generalized}. Furthermore, there exists a subset in the parameter space of higher-curvature couplings where these theories  only allow a massless spin-2 graviton propagating at the linearized level. 

In general, the field equations of this class contain  metric derivatives up to fourth order. However, for a static spherically symmetric (SSS) ansatz \eqref{sssansatz}, 
\begin{equation}\label{sssansatz}
   \mathrm{d}s^2=-f(r)\mathrm{d}t^2+\frac{1}{f(r)}\mathrm{d}r^2+r^2\mathrm{d}\Omega_{d-2, \kappa}^2,
\end{equation}
where $\mathrm{d}\Omega_{d-2, \kappa}^2$ describes the $(d-2)$-dimensional transverse surface of constant curvature normalized to $\kappa=+1,0,-1$ denoting spherical, flat and hyperbolic topologies, respectively. The metric function $f(r)$ is fully determined by the vanishing of a total derivative of a second order differential or algebraic expression, with the constant of integration related to the ADM mass. Theories with an algebraic equation of motion for $f$ are identified as quasi-topological (QT) gravities \cite{myers2010holographic,myers2010black,oliva2010new}, and they likely satisfy a Birkhoff theorem \cite{cisterna2017quintic,oliva2010new,oliva2011birkhoff,oliva2012birkhoff}. However, theories in the QT class only exist in $d\geq 5$. Another interesting GQT subclass  is that of Lovelock theories \cite{lovelock1970divergence,lovelock1971einstein}, which are the most direct generalizations of GR in that the field equations are always  
 second order differential equations for any metric. Similar to the QT class, for the ansatz \eqref{sssansatz},  the metric function $f$ is fully characterized by a single algebraic equation that only differs from that of the QT theories by an overall constant. Thus Lovelock gravity  corresponds to a subclass of QT theories. 
 
 However, Lovelock theories seem too restrictive -- Einstein gravity is the only possible Lovelock theory in $d=4$, and a Lovelock curvature density of order $n$ yields non-trivial dynamics only if $d>2n+1$. So the first non-trivial Lovelock theory appears in $d=5$, corresponding to  Gauss–Bonnet gravity with $n=2$, whereas non-trivial GQT gravity theories   exist for any order $n\geq3$ in $d\geq 4$ \cite{bueno2019generalized, Bueno:2022res}. In addition, as far as \eqref{sssansatz} is concerned,  GQT theories allow multiple inequivalent densities at a given order in $d\geq 5$, but QT (Lovelock) theories have one unique density at any order \cite{Bueno:2022res}. Thus GQTs constitute a much broader class of  higher-curvature theories than have already been specified.

We therefore choose to illustrate our approach in GQT gravity theories. 
The thermodynamics of these theories have not been explored 
to the same extent that Lovelock theories 
have.   Previous studies have been carried out in limited contexts, with only a few low order couplings in  dimensions 
not much larger than four \cite{ bueno2017black, bueno2017universal,  hennigar2017criticality, hennigar2017generalized, mir2019generalized,mir2019black}.  
Inspired by the multi-critical
behaviour 
found for a broad range of black holes in different contexts \cite{tavakoli2022multi, wu2022multicritical2, wu2022multicritical,wu2022thermodynamically}, our interests lie both 
in demonstrating our method and 
in understanding multi-criticality in GQT gravity. 

Our paper is organized as follows. In   sections \ref{sec:GQTthermo} and \ref{sec:constraints} we review some properties of GQT black hole solutions. In   section \ref{sec:Krule}, we provide an interpretation of multi-criticality and introduce the $K$-rule,  obtained by the reformulation of the Maxwell area rule. We develop in  section \ref{sec:construction}   a method based on the $K$-rule, carry out a discussion on its feasibility, and construct quadruple and quintuple points based on this approach.

\section{Fundamentals of thermodynamics of GQT black holes} \label{sec:GQTthermo}
Evaluated on a static spherically symmetric metric of the form \eqref{sssansatz}, the GQT class of curvature order $n\geq 2$ has exactly $n-1$ inequivalent densities in $d\geq 5$ \cite{Bueno:2022res}. As required by the integrability of the field equation for $f(r)$, on-shell GQT Lagrangian densities should be total derivatives of the form \cite{Bueno:2022res} 
\begin{equation}\label{gqtgd}
\mathcal{S}_n^{(k)}=\sum_{j=0}^{n}\lambda_{n,j}^{(k)}\mathcal{S}_{(n,j)},\quad k= 1,\cdots, k_n\equiv\mathrm{max}(1,n-1), \quad n\geq 1
\end{equation}
where $k$ labels one of the $n-1$ inequivalent densities, $\mathcal{S}_{(n,j)}$ is given by
\begin{equation}\label{totalD}   \mathcal{S}_{(n,j)}\equiv\frac{1}{r^{d-2}}\frac{\mathrm{d} }{\mathrm{d} r}\bigg[r^{d-1}\bigg(\frac{\kappa-f}{r^2}\bigg)^{n-j}\bigg(\frac{-f'}{2r}\bigg)^j\bigg],
\end{equation}
and $\lambda_{n,j}^{(k)}$ are some constrained coefficients such that $\mathcal{S}_{n}^{(k)}$ is induced by a real off-shell GQT density. Then the constraints
\begin{equation}\label{necessarycondition}
\sum_{j=0}^n\lambda_{n,j}^{(k)}(2n-dj)=0,\quad  \sum_{j=0}^n\lambda_{n,j}^{(k)}j(j-1)=0.
\end{equation}
are required to obtain the most general Lagrangian density \cite{Bueno:2022res}.  The first constraint ensures that all densities contribute with a power of $r^{d-1}$; the second ensures that the field equations are of 2nd order when linearized about constant curvature backgrounds. Note that any linear combination of on-shell densities still satisfies \eqref{gqtgd}, \eqref{totalD} and \eqref{necessarycondition}; therefore any density can be decomposed into $n-1$ independent densities. In other words, it is sufficient to study one particular choice of $\lambda_{n,j}^{(k)}$.

Incorporating  the constraints  \eqref{necessarycondition}, the following choice   
\begin{equation}\label{mygqtd}
    \lambda_{n,k+1}^{(k)}=(k-1)(dk-4n),\quad \lambda_{n,k}^{(k)}=2d(1-k^2)-4n(1-2k),\quad
    \lambda_{n,k-1}^{(k)}=k[d(k+1)-4n],
\end{equation}
of $\lambda_{n,j}^{(k)}$ form a family of GQT densities, 
with the remaining coefficients identical to $0$. This simple choice is employed in our analysis but the results in the remaining part of this section hold generally.

Upon integrating the equations of motion, we obtain \cite{bueno2019generalized} 
\begin{equation}\label{eom}
\frac{16\pi M}{(d-2)\Omega_{d-2,\kappa}}=\sum_{n=0}^{n_{\mathrm{max}}}\alpha_{n,k}\sum_{k=1}^{k_n}\sum_{j=0}^{n}\lambda_{n,j}^{(k)}\mathcal{F}_{(n,j)},
\end{equation}
where $\Omega_{d-2,\kappa}=\frac{2\pi^{\frac{d-1}{2}}}{\Gamma(\frac{d-1}{2})}$, $M$ is the ADM mass \cite{Deser:2002jk,Arnowitt:1960zzc,Arnowitt:1961zz,Arnowitt:1960es}, and 
\begin{align}
&\mathcal{F}_{(n,j)}=\frac{(-1)^{j+1}}{2^{j+1}}r^{d-2+j-2n}(\kappa-f)^{n-j-1}(f')^{j-2}\times 
\label{blockfeq}\\
&\{f'[j(d-1+j-2n)(\kappa-f)f-(j-1)r(\kappa+(n-j-1)f)f']+j(j-1)r(\kappa-f)ff''\} \nonumber
\end{align}
where prime is defined as taking derivatives with respect to $r$.

For $j=0, 1$,  $\mathcal{F}_{(n,j)}$ becomes an algebraic quantity (having no derivative terms), which implies the density defined by \eqref{mygqtd} with $k=1$ is purely QT (Lovelock). 
To make a consistent notation with common definitions, we separate QT densities from GQT ones \eqref{mygqtd} and define   
\begin{equation}\label{qtd}
    \mathcal{S}^{\mathrm{QT}}_{n}\equiv -\frac{1}{2}\mathcal{S}^{(1)}_{n}=+(2n-d)\mathcal{S}_{(n,0)}-2n\mathcal{S}_{(n,1)},
\end{equation}
as QT densities. The remainder,   with $k\geq2$, are  genuine GQT densities. 

Despite the complexity of \eqref{eom}, the thermodynamics of GQT black holes is determined by two simple equations:  
$f(r_+)=0$, which defines the outermost black hole horizon at $r=r_+$,  and $f'(r_+)=4\pi T$,
which defines the temperature of the black hole. 
These two relations are given by  
\cite{Bueno:2022res}
\begin{equation}
\begin{aligned}
    M&=\frac{\Omega_{d-2,\kappa}}{16\pi G}\sum_{n=0}^{n_{\mathrm{max}}}\sum_{k=1}^{k_n}\sum_{j=0}^{n}\alpha_{n,k}\lambda_{n,j}^{(k)}(j-1)\kappa^{n-j}r_+^{d-2n-1}(-2\pi r_+T)^j,
\end{aligned}
\end{equation}
\begin{equation}\label{eos}
\begin{aligned}
0=\sum_{n=0}^{n_{\mathrm{max}}}\sum_{k=1}^{k_n}\sum_{j=0}^{n}\alpha_{n,k}\lambda_{n,j}^{(k)}(d-2n+j-1)\kappa^{n-j}(-2\pi r_+T)^jr_+^{d-2n-2},
\end{aligned}
\end{equation}
where the couplings of the lowest two orders are set to be $ \alpha_{0,1}\equiv -(d-1)(d-2)/(2\ell^2)$, and $\alpha_{1,1}\equiv 1/2$ for   consistency with Einstein gravity. The parameter $\ell$ here is the AdS length, and so  $\alpha_{0,1}$ is identical to the cosmological constant $\Lambda$. In the context of black hole chemistry, all couplings except for $\alpha_{1,1}$ are identified as thermodynamic variables, with
\begin{equation}
    \Phi_{n}^{(k)}(r_+,T)\equiv\frac{\partial M(r_+,T,\{\alpha_{p,l}\})}{\partial \alpha_{n,k}}
\end{equation}
as their  corresponding  conjugate potentials.

As a well-known consequence in GR, the first law of thermodynamics and the Smarr relation hold as well in the extended phase space:
\begin{equation}\label{flaw}
\mathrm{d}M=T\mathrm{d}S+V\mathrm{d}P+\sum_{n\geq2}\Phi_{n}^{(k)}\mathrm{d}\alpha_{n,k},
\end{equation}
\begin{equation}\label{Smarr}
    (d-3)M=(d-2)TS-2PV+2\sum_{n\geq2}(n-1)\Phi_{n}^{(k)}\alpha_{n,k},
\end{equation}
where the pressure $P$ is a redefinition of $\alpha_{0,1}$ and $V$ is its corresponding conjugate volume 
\begin{equation}\label{P&V}
    P\equiv\frac{(d-1)(d-2)}{16\pi\ell^2},\quad V\equiv\frac{\Omega_{d-2,\kappa}r_+^{d-1}}{d-1}\; .
\end{equation}
The thermodynamic quantity $S$ is the Wald entropy \cite{Wald:1993nt}, which reads
\cite{Bueno:2022res}
\begin{equation}\label{entropy}
    S=\frac{\Omega_{d-2,\kappa}}{8G}\sum_{n=0}^{n_{\mathrm{max}}}\sum_{k=1}^{k_n}\sum_{j=0}^{n}\alpha_{n,k}\lambda_{n,j}^{(k)}j\kappa^{n-j}(-2\pi r_+T)^{j-1}r_+^{d-2n}.
\end{equation}
This quantity is not always positive, and in such situations it has been common  to simply discard solutions for which this is the case. However  ambiguities exist in the definition of the black hole entropy. For example,  adding to the Lagrangian  a term proportional to the induced metric on the horizon will, without having an effect on the other properties of the solution,  shift the entropy by an arbitrary constant.  One example is that of adding an  Euler density to the action \cite{mir2019black}. 
We shall therefore retain solutions with $S<0$ in considerations, appropriately indicating in our figures where this occurs.

Henceforth we shall consider the Gibbs free energy $G=M-TS$  for   investigating phase transitions.  The global minimum of $G$ yields the most stable thermodynamic phase at any given temperature.

Before continuing, we note that 
physical theories should only propagate one type of massless spin-2 graviton on constant curvature backgrounds. This in turn implies the effective Newtonian constant must have the same sign as the one in general relativity, which means that for the class of metrics   \eqref{sssansatz} having
asymptotically AdS solutions of the form 
\begin{equation}\label{farsol}
    f(r)=\kappa+f_{\infty}\frac{r^2}{\ell^2}+\frac{m}{h'(f_{\infty})r^{d-3}}+\cdots,
\end{equation} 
we shall only consider black holes with 
$f_{\infty}>0$, $  h'(f_{\infty})<0 $ and $\gamma^2>0$ \footnote{Those symbols are defined in appendix~\ref{sec:constraints}. } as satisfying the requisite physical criteria. We discuss these issues in
appendix~\ref{sec:constraints}.

\section{Geometric interpretation of interphase equilibrium }\label{sec:Krule}

We seek to obtain   the conditions under which three or more phases merge at a particular temperature and pressure.
The Gibbs free energy provides a diagnostic for this. Its global minimum as a function of the temperature $T$ 
 determines the thermodynamically stable state of the system for a given fixed choice of 
 the other thermodynamic parameters.   The presence of swallowtails in the Gibbs free energy indicates multiple phases, with first order phase transitions
 between two distinct phases taking place at the intersection point of the swallowtail. There must be $N-1$ swallowtails in order to have
 $N$ distinct phases.  Whenever the intersection points of $j$ different swallowtails coincide, then there is a $j$-th order multicritical point, where $j \leq N$.

Previous methods 
for finding multiple phases and $N$-tuple critical points exploited the fact that, regarding temperature as a function of horizon radius,  each extremum of  $T(r_+)$ corresponds to a cusp in the Gibbs free energy  \cite{tavakoli2022multi, wu2022multicritical2}. Hence
$N$ distinct phases 
require $2N-2$ distinct extrema.  If two adjacent inflections of $T(r_+)$ occur at the same temperature, then the intersection points of corresponding swallowtails will merge.  If this takes place for all the inflections, then all such intersection points will merge, correseponding to 
an $N$-tuple critical point. These critical points can be found by finely tuning the other thermodynamic parameters.

\begin{figure}[t]
\centering 
\includegraphics[width=.3\textwidth]{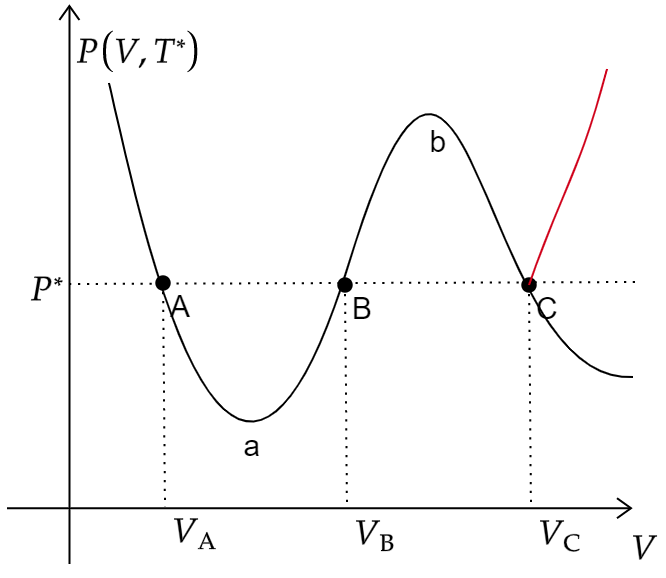}\includegraphics[width=.3\textwidth]{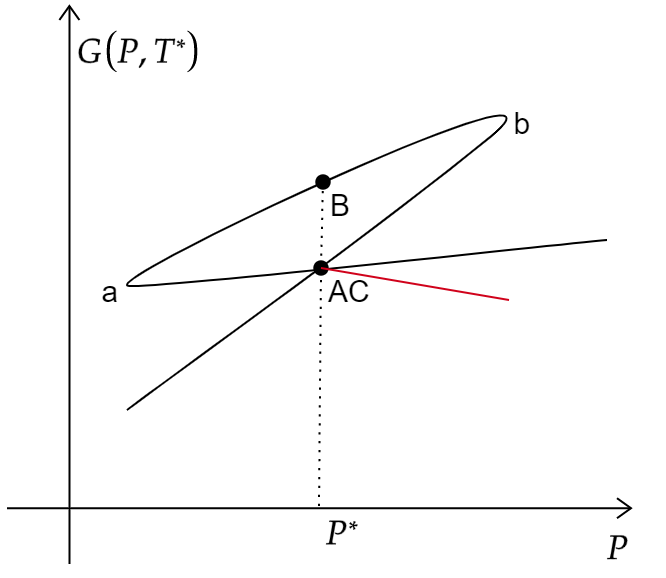}\includegraphics[width=.3\textwidth]{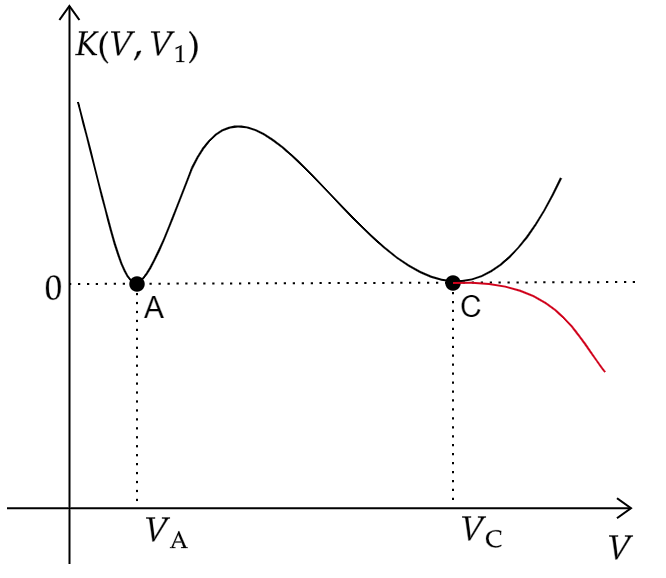}

\caption{\label{fig:PV&GP} The Maxwell equal-area construction implies $P=P^*$ divides AaBbC into two regions AaB and BbC with equal areas.  The red curve indicates the trajectories of the plots if $K'' = 0$.  } 
\end{figure}

Here we demonstrate an alternate method that is considerably more efficient. 
We start   with a brief review of the Maxwell construction \cite{clerk1875dynamical}. It is well-known that the multiplicity of the Gibbs free energy $G(P,T)$ corresponds to the non-monotonic behavior of the pressure $P(V,T)$. As illustrated in  figure \ref{fig:PV&GP}, a full oscillation AaBbC in the pressure at a fixed temperature $T^*$ leads to a swallowtail on the Gibbs phase diagram. With some abuse of notation, integrating $\mathrm{d}G$ along the loop $\mathrm{A}\rightarrow \mathrm{b} \rightarrow \mathrm{a}\rightarrow \mathrm{C}$ yields
\begin{equation}\label{equalarea1}
\begin{aligned}
    0=\oint \mathrm{d}G=\oint\frac{\partial G}{\partial P}\bigg|_{T^*}\mathrm{d}P=\int V\mathrm{d}P=PV\bigg|_{V_{\mathrm{A}}}^{V_{\mathrm{C}}}-\int P\mathrm{d}V=\int_{V_{\mathrm{A}}}^{V_{\mathrm{C}}} (P^*-P)\mathrm{d}V.
\end{aligned}
\end{equation}
The second equality holds because the temperature is fixed, the fourth one comes from integration by parts, and the last expression  follows from $P(V_{\mathrm{A}})=P(V_{\mathrm{C}})=P^*$, where $P^*$ characterizes the 
swallowtail intersection
point in the Gibbs phase plot. The geometric interpretation of \eqref{equalarea1} is obvious:  $P^*$ corresponds to a pressure that partitions 
the oscillatory parts of
the $P-V$ diagram into
equal areas. 

It is useful to define the  function $K(V,V_\mathrm{i})$ and its derivative $K'(V,V_\mathrm{i})$ as follows:
\begin{equation}
    K(V,V_\mathrm{i})\equiv \int_{V_\mathrm{i}}^{V} (P^*-P)\mathrm{d}V, \quad K'(V,V_\mathrm{i})\equiv\frac{\partial K}{\partial V}=P^*-P(V,T^*).
\end{equation}
It is obvious that 
$K(V_{\mathrm{A}},V_{\mathrm{A}})=0$ and that
the last expression of \eqref{equalarea1} can be rewritten as
\begin{equation}\label{example1}
    K(V_{\mathrm{C}},V_{\mathrm{A}})=0.
\end{equation}
However for any two points in thermodynamic phase space whose volumes satisfy \eqref{example1}, the relation \eqref{example1} alone doesn't imply that their difference in free energy   is zero. It is also necessary to 
ensure that $P(V_{\mathrm{C}}) = P(V_{\mathrm{A}})$ so that \eqref{equalarea1} holds. 
In a plot of $G$ vs. $P$,
this requirement is equivalent to the condition that A and C are the same point.
This in turn implies that 
\begin{equation}\label{example2}
    K'(V_{\mathrm{A}},V_{\mathrm{A}})=K'(V_{\mathrm{C}},V_{\mathrm{A}})=0
\end{equation}
where the first condition ensures that  $P(V_{\mathrm{A}},T^*)=P^*$.
Geometrically, \eqref{example1} and \eqref{example2} ensure that A and C are  the same point in the Gibbs energy diagram, and the continuity of $K$ between A and C
guarantees that this point is on some closed loop. A true self-intersection point  
(or double point) therefore emerges. 

We note that if the second derivative $K''$ 
\begin{equation}
    K'' \equiv\frac{\partial^2 K}{\partial V^2}=-\frac{\partial P}{\partial V}
\end{equation}
  vanishes at some point then  $K$
  will no longer be an extremum there. This is illustrated for point $C$ in the  rightmost diagram of  
  figure \ref{fig:PV&GP}  by the red curve.
The pressure will then be an extremum at this point
(as shown in the leftmost diagram in figure \ref{fig:PV&GP}), and the 
corresponding part of the curve in the free energy diagram will get reflected through $P^*$, as shown by the red curve in the middle diagram in figure \ref{fig:PV&GP}.
By convention,  we still regard this as a double point. 

These considerations can be easily generalized to any $N$-tuple point. We say that an $N$-tuple point exists at $(P^*,T^*)$ if and only if the $K$-rule is satisfied: namely that the function  $K(V,V_0)$ has $N$ real zero points $\{V_n\}$ for some fixed $V_0$ and $K'$ vanishes for all those roots, namely
\begin{equation}\label{Krule}
    K(V_n,V_0)=0, \quad K'(V_n,V_0)=0, \quad n=1,2,\cdots,N 
\end{equation}
are satisfied by exactly $N$ different values of $\{V_n\}$, including $V_0$ itself. 
Since the above argument about multicriticality is quite general, we would expect these discussions apply to any thermodynamic system for any conjugate pair of thermodynamic quantities, such as the temperature and the entropy. 

\section{Multiple phases and \texorpdfstring{$N$} --tuple critical points}\label{sec:construction}

We shall now construct multiple phases and $N$-tuple points 
for GQT black holes
based on the $K$-rule introduced in   section \ref{sec:Krule}. The procedure is simple.
\begin{enumerate}
    \item \label{1step} Write the function $K$ as 
\begin{equation}\label{myK}
    K(r_+,r_0)=\int_{r_0}^{r_+} \bigg[P^*-P(\tilde{r}_+,T^*,\{\alpha_{n,k}\})\bigg]\frac{\mathrm{d} V(\tilde{r}_+)}{\mathrm{d} \tilde{r}_+}\mathrm{d}\tilde{r}_+ ,
\end{equation}
where $P$ is given from  \eqref{eos} (which can be regarded as the equation of state)
with $\ell^2$ replaced by \eqref{P&V}, $V$ is identified as the thermodynamic volume defined in \eqref{P&V}, and $\{\alpha_{n,k}\}$ is the set of undetermined couplings.
    \item \label{2step} Apply the $K$-rule to $N$ positive distinct values of $r_+$ (where $r_0$ is taken to be any one of these values), then solve for $P^*, T^*, \{\alpha_{n,k}\}$ from the $2N-1$ independent equations\footnote{Minus one comes from the fact that $K(r_0,r_0)=0$ is trivial.} \eqref{Krule}. This implies that a minimum number of $2N-3$ non-zero higher-curvature couplings are required.
    \item \label{3step} Check if the solution $P^*, T^*, \{\alpha_{n,k}\}$ provides a real $N$-tuple point in a sense that exactly $N$ roots solve \eqref{Krule} as desired. If not, change the choice of as many horizon radii as needed until a real $N$-tuple point occurs. 
\end{enumerate}

We pause to make a few supplementary comments regarding the feasibility of the method.  First of all, $K$ is constructed from $P$ and $V$ instead of $T$ and $S$ because we want $K$ to be a simple function such that the equations in step \ref{2step} are solvable: the definition \eqref{myK} fulfills this  requirement since $K$ is in fact a polynomial in $r_+$. For convenience, $K$ is defined as a function of the radius rather than the volume.  It should be pointed out that the density $\mathcal{S}_{n\geq 3}^{(2)}$ is quasi-topological and becomes trivial in $d=2n$. Therefore, for even dimensions, we exclude $\alpha_{d/2,2}$ from our considerations. We shall also restrict ourselves to spherical black holes with $\kappa=1$ for simplicity. Since the pressure becomes a polynomial in $r_+$ (with the temperature $T$ considered as a non-dynamical parameter), then Descartes' rule of signs can be applied, which relates the largest number of oscillations in the region $r_+>0$ to the number of sign changes in the sequence of a polynomial's coefficients. Thus we discuss the feasibility of our method through studying the possibility for the occurrence of $N-1$ oscillations in $P$ by manipulating signs of couplings in the next paragraph.

 The feasibility of step \ref{2step} can be seen by induction. As indicated by \eqref{eos} or  table \ref{tab:pattern}, switching on an arbitrary   genuine GQT coupling ($k\geq 2$) always introduces three independent terms proportional to different powers of $r_+$ 
 in the expression for the  pressure in addition to 
\begin{equation}
    P_0=\frac{(d-2) T}{4 r_+}-\frac{(d-3) (d-2) \kappa }{16 \pi  r_+^2},
\end{equation}
which is the expression  with all higher-curvature couplings set to zero. Since $P_0$ has one sign change, then turning on any particular coupling can introduce another sign change. Respecting the fact that temperature and the coupling are free, it is possible to make $P$ have a full oscillation, which means that \eqref{Krule} has real solutions and a double point can be obtained. Similarly, for critical points involving more phases, we can obtain  additional oscillations by including two additional couplings per oscillation, as long as they switch on at least two monomials in $r_+$ that differ from those already present. Hence not any choice of $2N-3$ higher-order densities yields an equilibrium state with four phases or more. For example, as bolded in   table \ref{tab:pattern}, switching on $\{\alpha_{7,2},\alpha_{8,4},\alpha_{9,6},\alpha_{10,8},\alpha_{11,10}\}$ and keeping other couplings zero  only  introduces $1/r_+^{11}$, $1/r_+^{12}$, $1/r_+^{13}$ into the pressure.  Together with $P_0$, a total of five monomials are present in the expression for the pressure,  which is only enough to construct a triple point. In this example we must therefore   avoid turning on more than three couplings that contribute to the same three powers of $r_+$.

The number $2N-3$ should be considered as the minimum number of couplings required by our method. This may not be the smallest number of couplings needed for the emergence of an $N$-tuple point in general, since it is possible to have another sign change internally between the three monomials corresponding to the $\alpha_{n,k}$. The explicit form of new terms that are activated by switching on $\alpha_{n,k}$ reads
 \begin{equation}
 \begin{aligned}
     (-2)^{k-3}\alpha_{n,k}&\bigg\{(d + k - 2 n)(-1 + k) (d k - 4 n) \frac{\pi^kT^{k+1}\kappa^{n-k-1}}{r_+^{2n-k-1}} \\&+ (d+k-2n-1) [d (-1 + k^2) + (2 - 4 k) n]\frac{\pi^{k-1}T^{k}\kappa^{n-k}}{r_+^{2n-k}}\\&+(d + k - 2n-2)k(d + d k - 4 n) \frac{\pi^{k-2}T^{k-1}\kappa^{n-k+1}}{4r_+^{2n-k+1}}\bigg\}.
 \end{aligned}
 \end{equation}
We claim that there can be at most one sign change between these three terms. If the sign switches twice,    the first two coefficients must have a negative product, namely,
\begin{equation*}
    (d + k - 2 n)(-1 + k) (d k - 4 n)(d+k-2n-1) [d (-1 + k^2) + (2 - 4 k) n]<0.
\end{equation*}
Since we restrict ourselves to genuine GQT densities ($k\geq2$) in $d\geq 5$ only, the above can be simplified to 
\begin{equation}\label{uneq}
    (d + k - 2 n) (d+k-2n-1) (d k - 4 n) [d (-1 + k^2) + (2 - 4 k) n]<0.
\end{equation}
Note that the dimension $d$ cannot  be any of $2n-k,2n-k+1,2n-k+2$; otherwise one of the terms would vanish and it would be no longer possible to have two sign changes between two terms. Therefore the first two factors in \eqref{uneq} must be positive, which leads to  
\begin{equation*}
    (d k - 4 n) [d (-1 + k^2) + (2 - 4 k) n]<0,
\end{equation*}
finally giving 
\begin{equation}\label{myuneq1}
 \frac{dk}{4} <  n < \frac{d(k^2-1)}{4k-2},
\end{equation}
where $k\geq 2$ is applied to determine the directions of inequalities. Meanwhile, we require the last two terms produce a sign change as well. Through a  similar analysis, we arrive at 
\begin{equation*}
   [d (-1 + k^2) + (2 - 4 k) n] (d + d k - 4 n)<0,
\end{equation*}
which yields 
\begin{equation}\label{myuneq2}
    \frac{d(k^2-1)}{4k-2}< n < \frac{d(k+1)}{4}.
\end{equation}

Our claim is thus proved, since \eqref{myuneq2} contradicts   \eqref{myuneq1}. Even if there is an extra internal sign change,  a half oscillation in the pressure does not necessarily occur, since these variables   are not only integers but also constrained relative to each other in a complicated way. However, notice that the range of $n$ in \eqref{myuneq2} is proportional to $d$, implying that we have a rather large parameter space for  $n$ and $k$. Hence we can still expect that there exists some choices of $d,k,n$ that can give rise to an $N$-tuple point with less than $2N-3$ couplings. As a consequence, we would also expect $N-1$ to be a lower bound on the number of couplings needed for the occurrence of an $N$-tuple point. Because of these considerations,  step \ref{3step} is added to guarantee that exactly $N$ phases are obtained.

From the previous discussion, we can see that to figure out a general Gibbs phase rule is challenging. For neutral multi-rotating black holes \cite{wu2022multicritical}, because the phase structure is invariant under the exchange of any two angular momenta, turning on any two additional couplings always creates a new phase and vice versa. However in the GQT scenario, this symmetry is broken between any two couplings, and theories with distinct values of $k$ (even if they have the same $n$) differ a lot in their phase structures. A multi-critical point may not occur even if infinitely many couplings are turned on. This implies that the problem is not as simple as it is in  Lovelock gravity where $k=1$ and each density \eqref{qtd}  similarly contributes to the thermodynamics \cite{wu2022multicritical2}. We leave this question for future investigation.

\begin{sidewaysfigure}
\centering 
\includegraphics[width=1\textwidth]{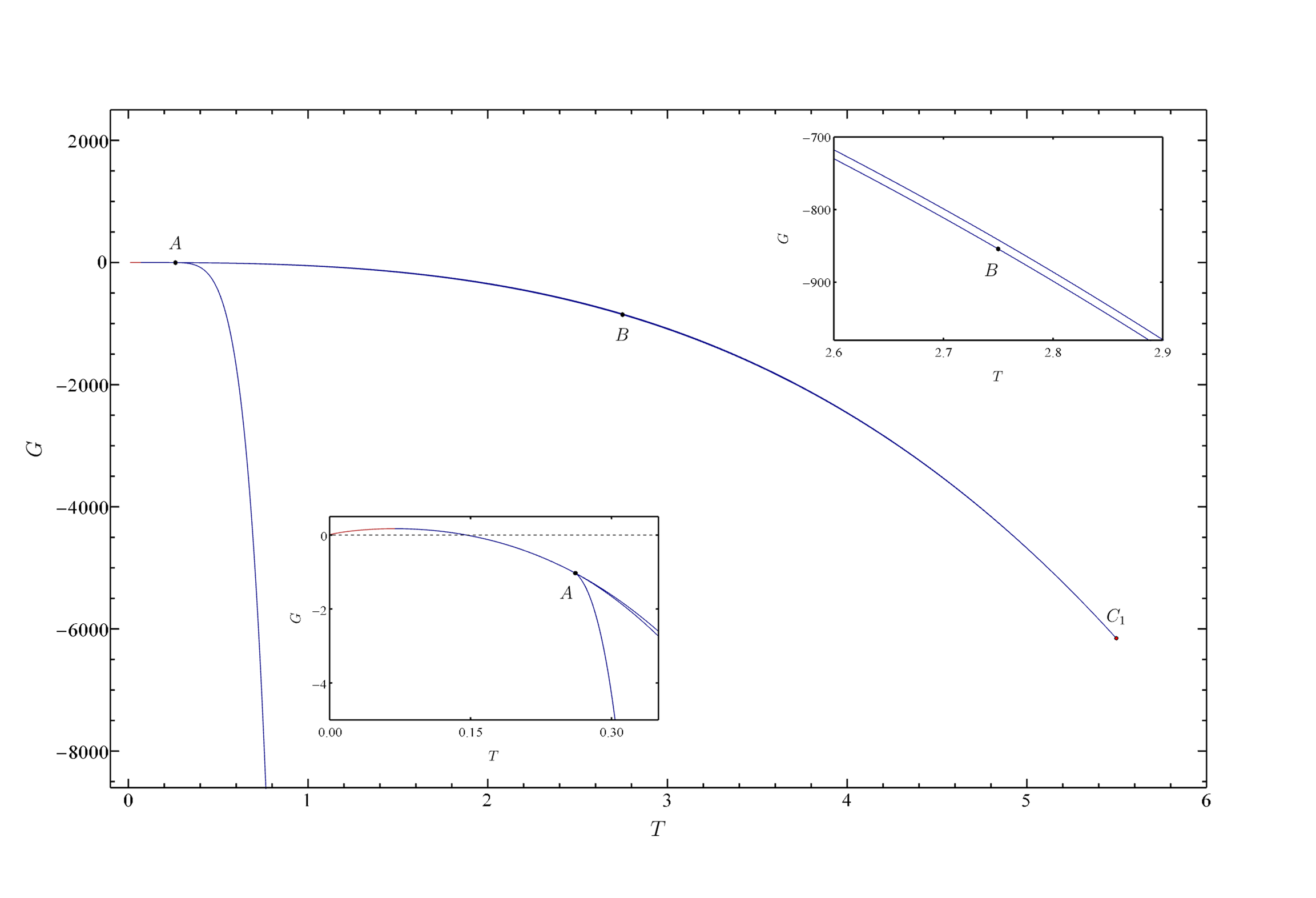}
\caption{\label{fig:4tuplefig1}  A quadruple point $A$ is constructed in $d=7$ with 
$f_{\infty}=1.001359562$, $P^*=0.07466400248$, $T^*=0.2615575508$,  $\alpha_{4,2}=-1.983132445$, $\alpha_{5,2}=-0.6595495180$, $\alpha_{6,2}=0.1064022549$,  $\alpha_{7,2}=-0.02347725421$, $\alpha_{8,2}=0.002665217984$. Four exact radii $\{1,1.07,1.239,1.4\}$ are taken as input. The black hole has positive $\gamma^2$ everywhere while the entropy is almost positive everywhere except the red region near $T=0$. Behaviors around $A$ and $B$ are indicated by the two subfigures respectively. The point $C_1$ is one of the cusps, and in the next figure \ref{fig:4tuplefig2}, we show the rest $5$ which are accumulated around $A$. So it is a quadruple point with an extremely far-away cusp. 
}

\end{sidewaysfigure}

\begin{sidewaysfigure}
\centering 
\includegraphics[width=1\textwidth]{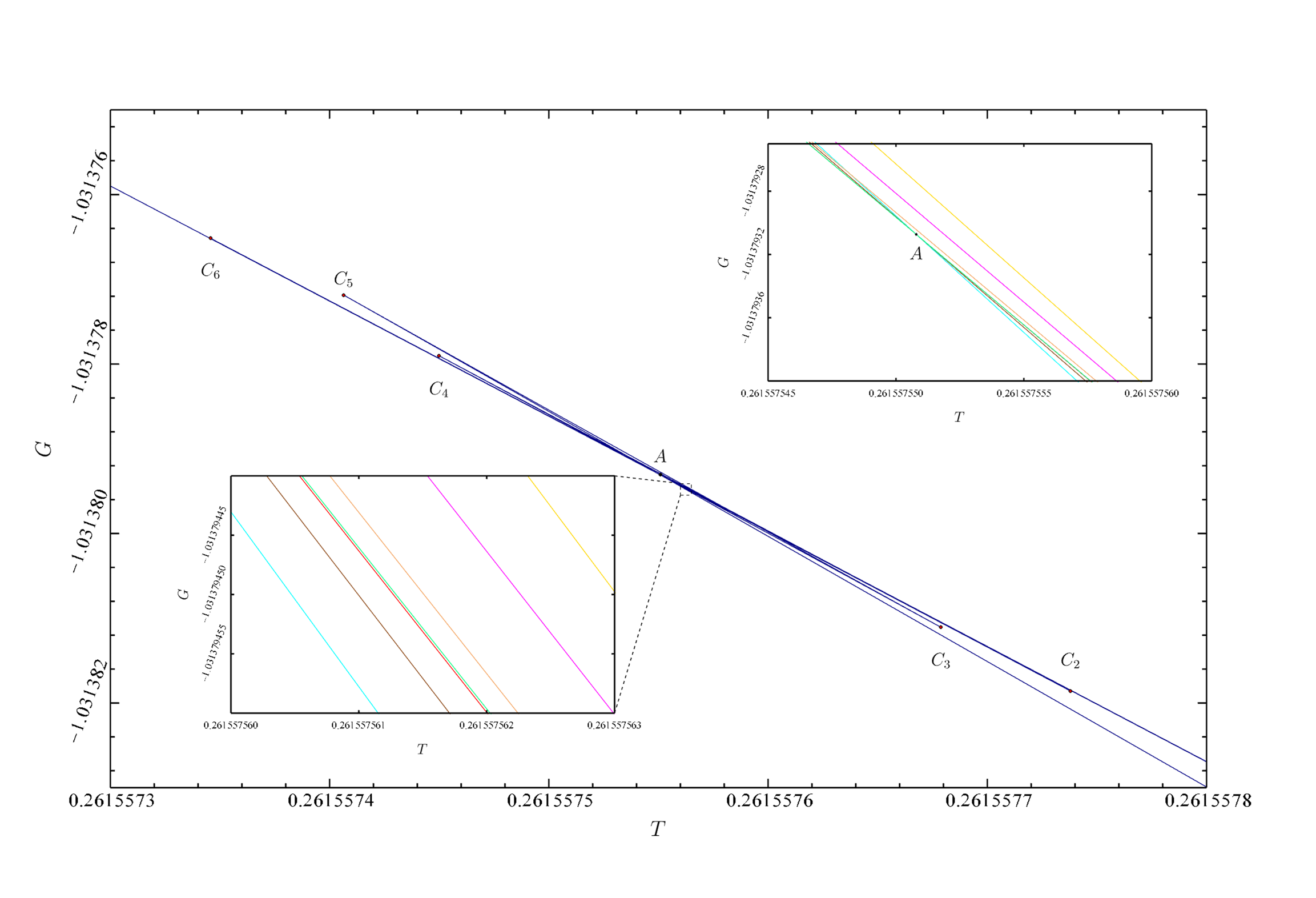}
\caption{\label{fig:4tuplefig2}  The figure shows further magnifications about the quadruple point $A$ in the figure \ref{fig:4tuplefig1}. The points $C_2$ to $C_6$ are five cusps, and the rest one is indicated in the figure \ref{fig:4tuplefig1}. These cusps are extremely narrow so that they look like lines. Regarding this fact, we further zoom around $A$ and color curves passing $A$ for better distinguishability. As getting closer to $A$, those curves are closer to each other. If staring at the right part of the right subfigure carefully enough, we can see there are three curves pass $A$,  but it is actually four, because the green and red curves between orange and brown ones overlap. Moving a bit far away from $A$, namely in the dotted box, the red and green curves become more distinguishable.  
}

\end{sidewaysfigure}

\begin{sidewaysfigure}
\centering 
\includegraphics[width=1\textwidth]{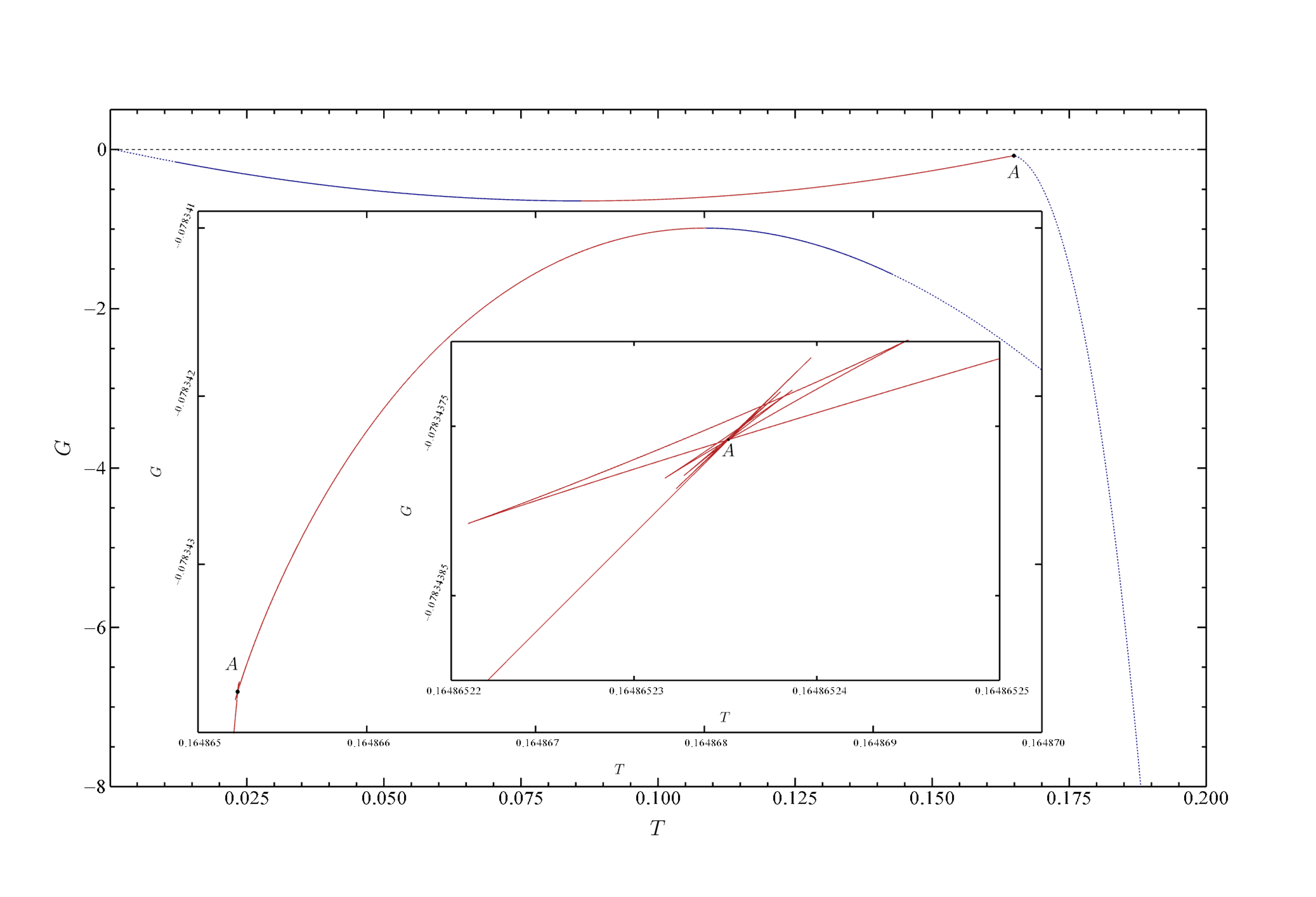}
\caption{\label{fig:5tuplefig1}   A quintuple point $A$ appears in $d=7$ with $f_{\infty}=0.9982618066$, $P^*=0.03402189695$, $T^*=0.1648652352$,  $\alpha_{3,2}=1$,$\alpha_{4,2}=6.232655784$, $\alpha_{5,2}=1.096317618$, $\alpha_{6,2}=0.2676712136$,  $\alpha_{7,2}=-0.2079188594$, $\alpha_{8,2}=0.09653639807$, $\alpha_{9,2}=-0.02640506854$,$\alpha_{10,2}=0.003268616550$, where $\alpha_{3,2}$ is set to $1$ in advance, and the rest of parameters are computed after five exact radii $\{1,1.07,1.21,1.4,1.6\}$ are taken as input. The upside of this adjustment is that we can find a physical quintuple point with negative free energy more easily. Blue and red colors indicate regions with positive and negative entropy respectively; solid and dashed curves partition the points with positive and negative $\gamma^2$ respectively. Two subfigures show different extents of magnification about the quintuple point $A$ and a further magnification is shown in the next figure \ref{fig:5tuplefig2} to clarify $A$ is truly a quintuple point.
} 

\end{sidewaysfigure}

\begin{sidewaysfigure}
\centering 
\includegraphics[width=1\textwidth]{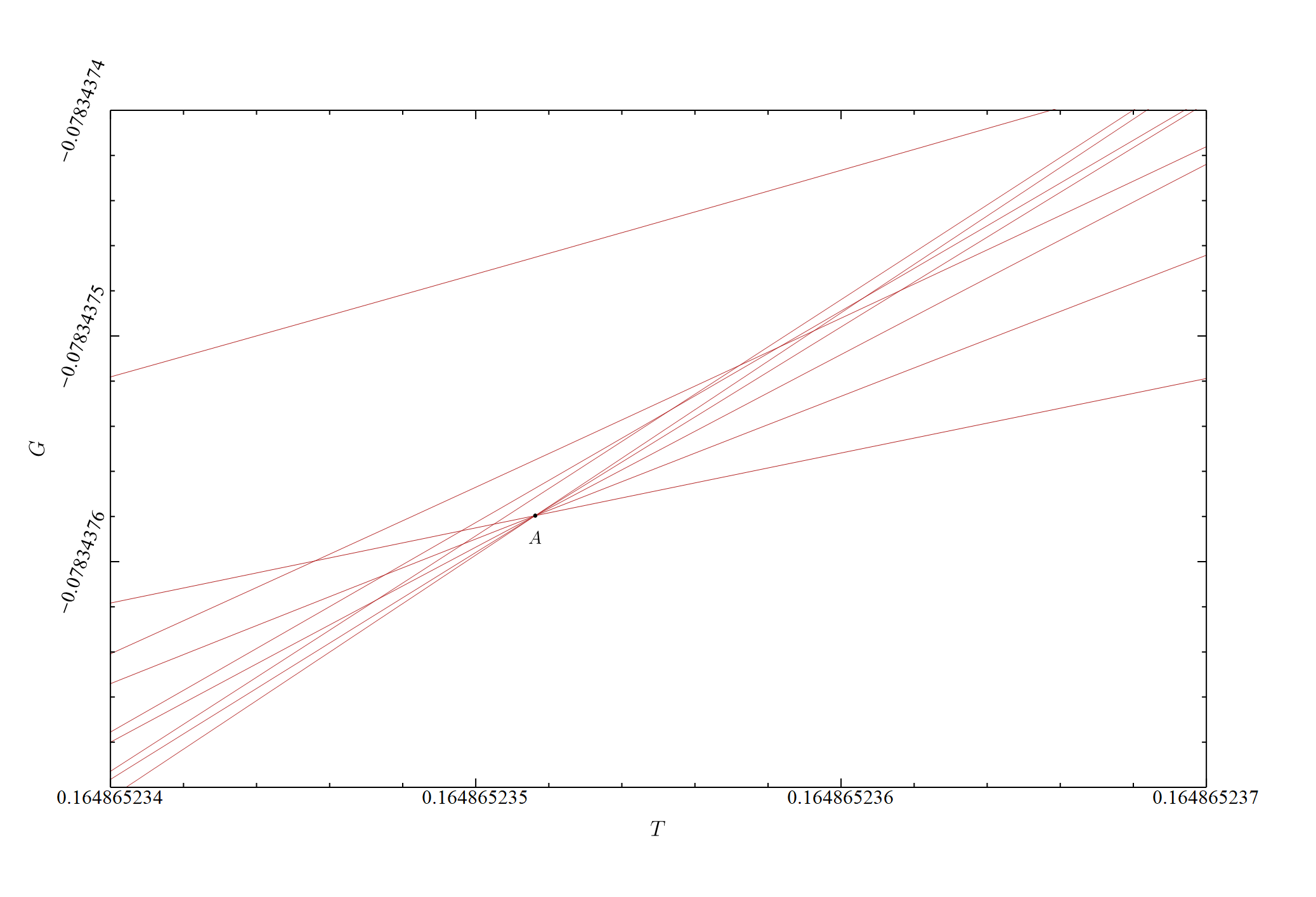}
\caption{\label{fig:5tuplefig2}   The figure shows further magnifications about the quintuple point $A$ in the figure \ref{fig:5tuplefig1}. We can clearly see the intersection of $5$ curves appears at $A$. 
}

\end{sidewaysfigure}

In order to obtain multiple phases and multi-critical points, the physical constraints and ensuring $P>0$ everywhere must also be considered. Since it is difficult to find a case with positive $\gamma^2$ everywhere, we only impose  $\gamma^2 > 0$   in a neighbourhood of each critical point. In practice, we keep manipulating $r_+$'s until a critical point satisfying all constraints occurs. Under these considerations, we  explicitly obtain a quadruple point (figures \ref{fig:4tuplefig1} and \ref{fig:4tuplefig2}) and a quintuple point (figures \ref{fig:5tuplefig1} and \ref{fig:5tuplefig2}) for two different  spherical GQT black holes.
Note that to see the merging of multiple swallowtails requires high precision in computations due to the finely tuned nature of multi-critical points.
Both multi-critical points have negative  Gibbs free energies,   implying stable phase transitions. For the quintuple point, an extra coupling $\alpha_{3,2}$ is fixed to be $1$ before running the procedure in order to make physical cases easier to emerge.

Compared with the previous methods, the advantages of our procedure are quite significant. First, the application of the Maxwell construction turns the problem into algebra which enables our method to produce critical points with arbitrarily high precision such that very tiny phase structures can be discovered easily. More importantly, the method is more efficient in the sense that it does not need any fine-tuning procedure such as those previously employed \cite{tavakoli2022multi,wu2022multicritical,wu2022multicritical2,wu2022thermodynamically} where $N$-tuple points were obtained by manipulating thermodynamic variables so that a common point of inflection occurred between multiple maxima and minima of the temperature as a function of $r_+$. 
However, due to the additional physical constraint $\gamma^2>0$ induced from the non-algebraic nature of equation of motions in genuine GQTGs, finding a physical multicritical point with a large $N$ is still time-consuming.

\section{Conclusions}

We have exploited   Maxwell's equal area law  to find an interphase equilibrium for black holes with multiple phases, reformulating it into what we call the $K$-rule.
 Utilizing the $K$-rule, we developed a novel approach for constructing $N$-tuple points in the phase space of black holes.
 
 We applied our results
 to GQT theories 
 with $2N-3$ genuine  couplings. Our analysis suggests that the minimum number of couplings required for the formation of an $N$-tuple point is likely confined between $N-1$ and $2N-3$. We presented quadruple and quintuple points to illustrate the effectiveness of the method.

The discovery that black holes can have multiple phases and multicritical behaviour has interesting implications for quantum gravity.  The distinct phases correspond to different thermodynamically stable states having different entropy, analogous to the way in which steam and ice are respectively  high-entropy and low-entropy states of water.
Systems having multicritical behaviour, such as colloids and polymers
\cite{peters2020defying, garcia2018depletion}, interact not only through short-range hard-sphere (or hard-cylinder) effects but also through
additional soft or long-range interactions comprising multiple length sales \cite{peters2020defying}.  It is reasonable to infer that the fundamental degrees of freedom of a black hole likewise have such complicated interactions, commensurate with recent indications that they are molecular in character \cite{Wei:2019uqg,Wei:2019yvs}.

 Future work would involve applying the $K$-rule to other kinds of black hole holes, particularly those whose horizon structures are not spherically symmetric.  These include accelerating black holes
 in non-linear electrodynamics 
 \cite{Hale:2023qjx}, multiply rotating black holes \cite{wu2022thermodynamically}, 
 and various black hole solutions in supergravity theories \cite{Caceres:2015vsa}. 


\begin{sidewaystable}
\centering
\begin{tabular}{llllllllllllllllll} 
\cline{1-3}
\multicolumn{1}{|l}{3}              & 4                     & \multicolumn{1}{l|}{5}              &                       &                                     &                       &                                     &                       &                                     &                       &                                     &                  &                                     &                  &                                     &                 &                                      &                                                     \\ 
\cline{1-5}
\multicolumn{1}{|l}{\myb{$\alpha_{3,2}$}} & \myb{$\alpha_{3,2}$}        & \multicolumn{1}{l|}{\myb{$\alpha_{3,2}$}} & 6                     & \multicolumn{1}{l|}{7}              &                       &                                     &                       &                                     &                       &                                     &                  &                                     &                  &                                     &                 &                                      &                                                     \\ 
\cline{1-2}\cline{4-7}
                                    & \multicolumn{1}{l|}{} & \myb{$\alpha_{4,2}$}                      & \myb{$\alpha_{4,2}$}        & \multicolumn{1}{l|}{\myb{$\alpha_{4,2}$}} & 8                     & \multicolumn{1}{l|}{9}              &                       &                                     &                       &                                     &                  &                                     &                  &                                     &                 &                                      &                                                     \\ 
\cline{2-2}\cline{6-9}
\multicolumn{1}{l|}{}               & \myv{$\alpha_{4,3}$}        & \myv{$\alpha_{4,3}$}                      & \myv{$\alpha_{4,3}$}        & \myb{$\alpha_{5,2}$}                      & \myb{$\alpha_{5,2}$}        & \multicolumn{1}{l|}{\myb{$\alpha_{5,2}$}} & 10                    & \multicolumn{1}{l|}{11}             &                       &                                     &                  &                                     &                  &                                     &                 &                                      &                                                     \\ 
\cline{2-3}\cline{8-11}
                                    &                       & \multicolumn{1}{l|}{}               & \myv{$\alpha_{5,3}$}        & \myv{$\alpha_{5,3}$}                      & \myv{$\alpha_{5,3}$}        & \myb{$\alpha_{6,2}$}                      & \myb{$\alpha_{6,2}$}        & \multicolumn{1}{l|}{\myb{$\alpha_{6,2}$}} & 12                    & \multicolumn{1}{l|}{13}             &                  &                                     &                  &                                     &                 &                                      &                                                     \\ 
\cline{3-3}\cline{10-13}
                                    & \multicolumn{1}{l|}{} & \myr{$\alpha_{5,4}$}                      & \myr{$\alpha_{5,4}$}        & \myr{$\alpha_{5,4}$}                      & \myv{$\alpha_{6,3}$}        & \myv{$\alpha_{6,3}$}                      & \myv{$\alpha_{6,3}$}        & \boldmath\myb{$\alpha_{7,2}$}                      & \boldmath\myb{$\alpha_{7,2}$}        & \multicolumn{1}{l|}{\boldmath\myb{$\alpha_{7,2}$}} & 14               & \multicolumn{1}{l|}{15}             &                  &                                     &                 &                                      &                                                     \\ 
\cline{3-4}\cline{12-15}
                                    &                       &                                     & \multicolumn{1}{l|}{} & \myr{$\alpha_{6,4}$}                      & \myr{$\alpha_{6,4}$}        & \myr{$\alpha_{6,4}$}                      & \myv{$\alpha_{7,3}$}        & \myv{$\alpha_{7,3}$}                      & \myv{$\alpha_{7,3}$}        & \myb{$\alpha_{8,2}$}                      & \myb{$\alpha_{8,2}$}   & \multicolumn{1}{l|}{\myb{$\alpha_{8,2}$}} & 16               & \multicolumn{1}{l|}{17}             &                 &                                      &                                                     \\ 
\cline{4-4}\cline{14-17}
                                    &                       & \multicolumn{1}{l|}{}               & \myt{$\alpha_{6,5}$}        & \myt{$\alpha_{6,5}$}                      & \myt{$\alpha_{6,5}$}        & \myr{$\alpha_{7,4}$}                      & \myr{$\alpha_{7,4}$}        & \myr{$\alpha_{7,4}$}                      & \myv{$\alpha_{8,3}$}        & \myv{$\alpha_{8,3}$}                      & \myv{$\alpha_{8,3}$}   & \myb{$\alpha_{9,2}$}                      & \myb{$\alpha_{9,2}$}   & \multicolumn{1}{l|}{\myb{$\alpha_{9,2}$}} & 18              & \multicolumn{1}{l|}{19}              &                                                     \\ 
\cline{4-5}\cline{16-18}
                                    &                       &                                     &                       & \multicolumn{1}{l|}{}               & \myt{$\alpha_{7,5}$}        & \myt{$\alpha_{7,5}$}                      & \myt{$\alpha_{7,5}$}        & \boldmath\myr{$\alpha_{8,4}$}                      & \boldmath\myr{$\alpha_{8,4}$}        & \boldmath\myr{$\alpha_{8,4}$}                      & \myv{$\alpha_{9,3}$}   & \myv{$\alpha_{9,3}$}                      & \myv{$\alpha_{9,3}$}   & \myb{$\alpha_{10,2}$}                     & \myb{$\alpha_{10,2}$} & \multicolumn{1}{l|}{\myb{$\alpha_{10,2}$}} & \begin{tabular}[c]{@{}l@{}}$\cdots$\\\end{tabular}  \\ 
\cline{5-5}\cline{18-18}
                                    &                       &                                     & \multicolumn{1}{l|}{} & \myc{$\alpha_{7,6}$}                      & \myc{$\alpha_{7,6}$}        & \myc{$\alpha_{7,6}$}                      & \myt{$\alpha_{8,5}$}        & \myt{$\alpha_{8,5}$}                      & \myt{$\alpha_{8,5}$}        & \myr{$\alpha_{9,4}$}                      & \myr{$\alpha_{9,4}$}   & \myr{$\alpha_{9,4}$}                      & \myv{$\alpha_{10,3}$}  & \myv{$\alpha_{10,3}$}                     & \myv{$\alpha_{10,3}$} & \myb{$\alpha_{11,2}$}                      & $\cdots$                                            \\ 
\cline{5-6}
                                    &                       &                                     &                       &                                     & \multicolumn{1}{l|}{} & \myc{$\alpha_{8,6}$}                      & \myc{$\alpha_{8,6}$}        & \myc{$\alpha_{8,6}$}                      & \myt{$\alpha_{9,5}$}        & \myt{$\alpha_{9,5}$}                      & \myt{$\alpha_{9,5}$}   & \myr{$\alpha_{10,4}$}                     & \myr{$\alpha_{10,4}$}  & \myr{$\alpha_{10,4}$}                     & \myv{$\alpha_{11,3}$} & \myv{$\alpha_{11,3}$}                      & $\cdots$                                            \\ 
\cline{6-6}
                                    &                       &                                     &                       & \multicolumn{1}{l|}{}               & \myo{$\alpha_{8,7}$}        & \myo{$\alpha_{8,7}$}                      & \myo{$\alpha_{8,7}$}        & \boldmath\myc{$\alpha_{9,6}$}                      & \boldmath\myc{$\alpha_{9,6}$}        & \boldmath\myc{$\alpha_{9,6}$}                      & \myt{$\alpha_{10,5}$}  & \myt{$\alpha_{10,5}$}                     & \myt{$\alpha_{10,5}$}  & \myr{$\alpha_{11,4}$}                     & \myr{$\alpha_{11,4}$} & \myr{$\alpha_{11,4}$}                      & $\cdots$                                            \\ 
\cline{6-7}
                                    &                       &                                     &                       &                                     &                       & \multicolumn{1}{l|}{}               & \myo{$\alpha_{9,7}$}        & \myo{$\alpha_{9,7}$}                      & \myo{$\alpha_{9,7}$}        & \myc{$\alpha_{10,6}$}                     & \myc{$\alpha_{10,6}$}  & \myc{$\alpha_{10,6}$}                     & \myt{$\alpha_{11,5}$}  & \myt{$\alpha_{11,5}$}                     & \myt{$\alpha_{11,5}$} & \myr{$\alpha_{12,4}$}                      & $\cdots$                                            \\ 
\cline{7-7}
                                    &                       &                                     &                       &                                     & \multicolumn{1}{l|}{} & \myor{$\alpha_{9,8}$}                      & \myor{$\alpha_{9,8}$}        & \myor{$\alpha_{9,8}$}                      & $\myo{\alpha_{10,7}}$       & $\myo{\alpha_{10,7}}$                     & $\myo{\alpha_{10,7}}$  & \myc{$\alpha_{11,6}$}                     & \myc{$\alpha_{11,6}$}  & \myc{$\alpha_{11,6}$}                     & \myt{$\alpha_{12,5}$} & \myt{$\alpha_{12,5}$}                      & $\cdots$                                            \\ 
\cline{7-8}
                                    &                       &                                     &                       &                                     &                       &                                     & \multicolumn{1}{l|}{} & \boldmath\myor{$\alpha_{10,8}$}                     & \boldmath\myor{$\alpha_{10,8}$}       & \boldmath\myor{$\alpha_{10,8}$}                     & $\myo{\alpha_{11,7}}$  & $\myo{\alpha_{11,7}}$                     & $\myo{\alpha_{11,7}}$  & \myc{$\alpha_{12,6}$}                     & \myc{$\alpha_{12,6}$} & \myc{$\alpha_{12,6}$}                      & $\cdots$                                            \\ 
\cline{8-8}
                                    &                       &                                     &                       &                                     &                       & \multicolumn{1}{l|}{}               & \mybl{$\alpha_{10,9}$}       & \mybl{$\alpha_{10,9}$}                     & \mybl{$\alpha_{10,9}$}       & \myor{$\alpha_{11,8}$}                     & \myor{$\alpha_{11,8}$}  & \myor{$\alpha_{11,8}$}                     & $\myo{\alpha_{12,7}}$  & $\myo{\alpha_{12,7}}$                     & $\myo{\alpha_{12,7}}$ & $\myc{\alpha_{13,6}}$                      & $\cdots$                                            \\ 
\cline{8-9}
                                    &                       &                                     &                       &                                     &                       &                                     &                       & \multicolumn{1}{l|}{}               & \mybl{$\alpha_{11,9}$}       & \mybl{$\alpha_{11,9}$}                     & \mybl{$\alpha_{11,9}$}  & \myor{$\alpha_{12,8}$}                     & \myor{$\alpha_{12,8}$}  & \myor{$\alpha_{12,8}$}                     & $\myo{\alpha_{13,7}}$ & $\myo{\alpha_{13,7}}$                      & $\cdots$                                            \\ 
\cline{9-9}
                                    &                       &                                     &                       &                                     &                       &                                     & \multicolumn{1}{l|}{} & \boldmath\myg{$\alpha_{11,10}$}                    & \boldmath\myg{$\alpha_{11,10}$}      & \boldmath\myg{$\alpha_{11,10}$}                    & \mybl{$\alpha_{12,9}$}  & \mybl{$\alpha_{12,9}$}                     & \mybl{$\alpha_{12,9}$}  & \myor{$\alpha_{13,8}$}                     & \myor{$\alpha_{13,8}$} & \myor{$\alpha_{13,8}$}                      & $\cdots$                                            \\ 
\cline{9-10}
                                    &                       &                                     &                       &                                     &                       &                                     &                       &                                     & \multicolumn{1}{l|}{} & \myg{$\alpha_{12,10}$}                    & \myg{$\alpha_{12,10}$} & \myg{$\alpha_{12,10}$}                    & \mybl{$\alpha_{13,9}$}  & \mybl{$\alpha_{13,9}$}                     & \mybl{$\alpha_{13,9}$} & \myor{$\alpha_{14,9}$}                      & $\cdots$                                            \\ 
\cline{10-10}
                                    &                       &                                     &                       &                                     &                       &                                     &                       & \multicolumn{1}{l|}{}               & \myp{$\alpha_{12,11}$}      & \myp{$\alpha_{12,11}$}                    & \myp{$\alpha_{12,11}$} & \myg{$\alpha_{13,10}$}                    & \myg{$\alpha_{13,10}$} & \myg{$\alpha_{13,10}$}                    & \mybl{$\alpha_{14,9}$} & \mybl{$\alpha_{14,9}$}                      & $\cdots$                                            \\ 
\cline{10-11}
                                    &                       &                                     &                       &                                     &                       &                                     &                       &                                     &                       &                                     & $\vdots$         & $\vdots$                            & $\vdots$         & $\vdots$                            & $\vdots$        & $\vdots$                             & $\ddots$                                           
\end{tabular}
\caption{\label{tab:pattern} The table shows a general pattern that the pressure follows when some genuine GQT densities ($k\geq 2$) are turned on. The top element of each column indicates the power of $1/r_+$, and each row contains all possible couplings with a constant sum of subscripts. The table tells what powers of $1/r_+$ in the expression of pressure are influenced by which couplings. For example, if $\alpha_{5,2}$ is the only non-zero coupling, then the pressure will contain three monomials of the form $1/r_+^7, 1/r_+^8, 1/r_+^9$ in additional to $P_0$; if $\alpha_{7,4}$ is turned on as well, then two extra monomials $1/r_+^{10},1/r_+^{11}$ will be introduced and both of $\alpha_{5,2}$ and $\alpha_{7,4}$ contribute to coefficient of the term $1/r_+^9$. }
\end{sidewaystable}

\clearpage

\clearpage

\clearpage

\clearpage

\clearpage

\begin{sidewaysfigure}
\centering 
\includegraphics[width=1\textwidth]{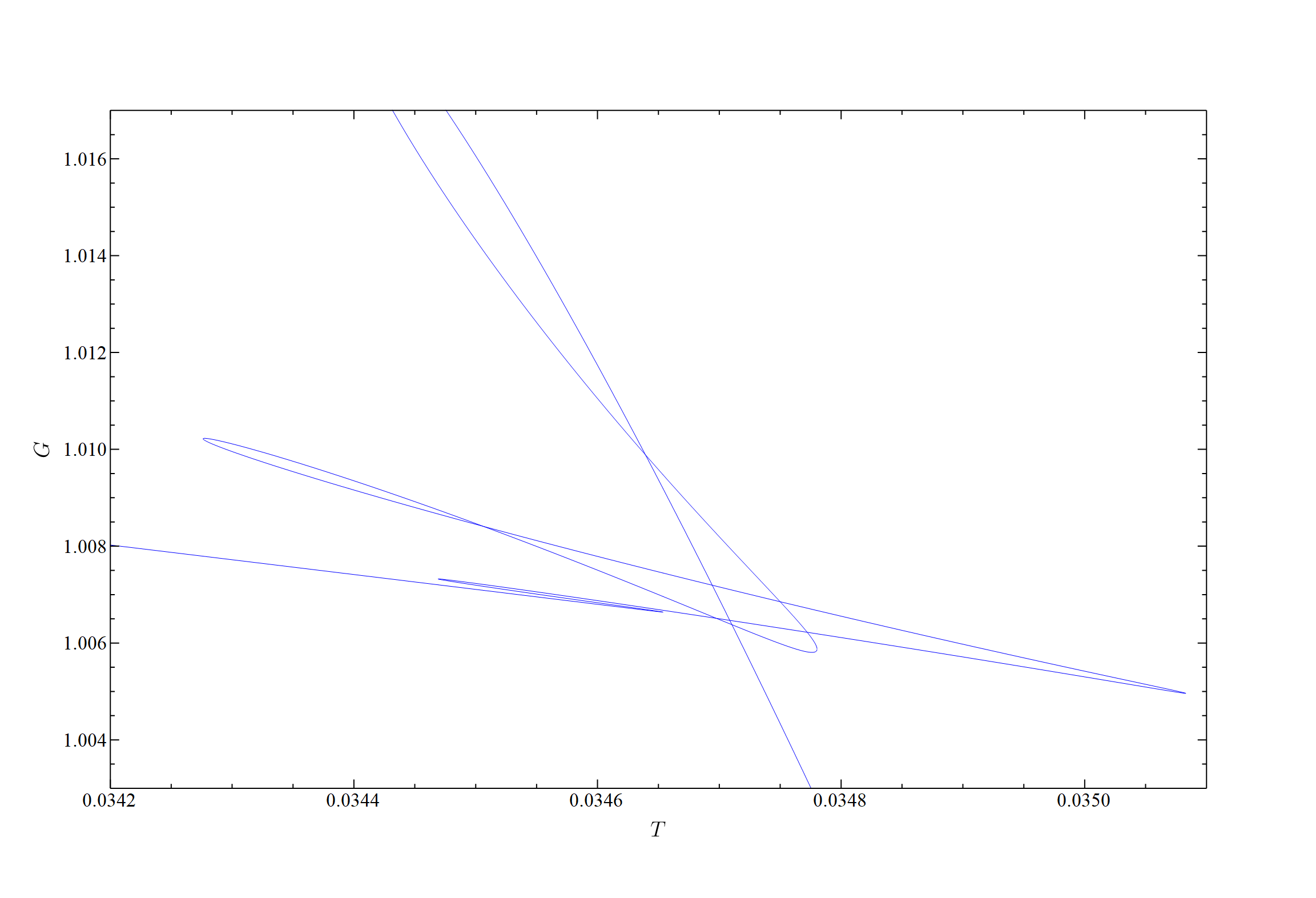}
\caption{\label{fig:PMW0}   The figure shows the Gibbs free energy corresponds to the equation of states \eqref{tempEPM} with $Q=1.2368, P^*=0.0020878, b_5=-1.0973,b_9=1.5937,b_{13}=-1.2746,b_{17}=0.40172$ which intersects with $T^*=0.034711$  at $r_+^{(1)}=1,r_+^{(2)}=1.04,r_+^{(3)}=1.1,r_+^{(4)}=1.4,r_+^{(5)}=1.9,r_+^{(6)}=2.9,r_+^{(7)}=4$. 
}

\end{sidewaysfigure}

\clearpage

\begin{sidewaysfigure}
\centering 
\includegraphics[width=1\textwidth]{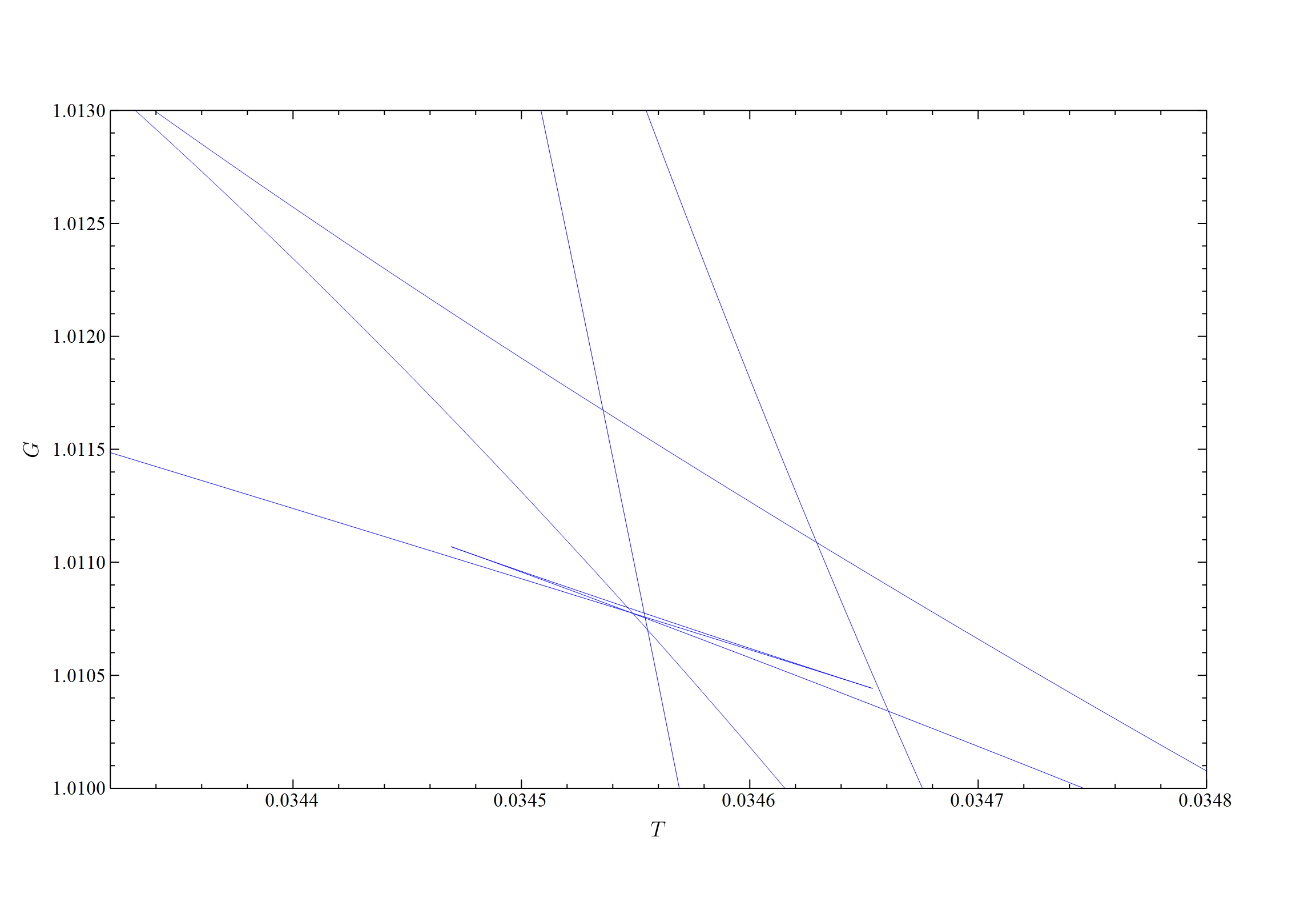}
\caption{\label{fig:PMW1}   The figure shows the Gibbs free energy corresponds to the equation of states \eqref{tempEPM} with $Q=1.2424, P^*=0.0020704, b_5=-1.1233,b_9=1.6535,b_{13}=-1.3338,b_{17}=0.42279$ 
which intersects with $T^*=0.034555$ at $r_+^{(1)}=1,r_+^{(2)}=1.04,r_+^{(3)}=1.1,r_+^{(4)}=1.42,r_+^{(5)}=1.9,r_+^{(6)}=2.93,r_+^{(7)}=4$.
}

\end{sidewaysfigure}

\clearpage

\begin{sidewaysfigure}
\centering 
\includegraphics[width=1\textwidth]{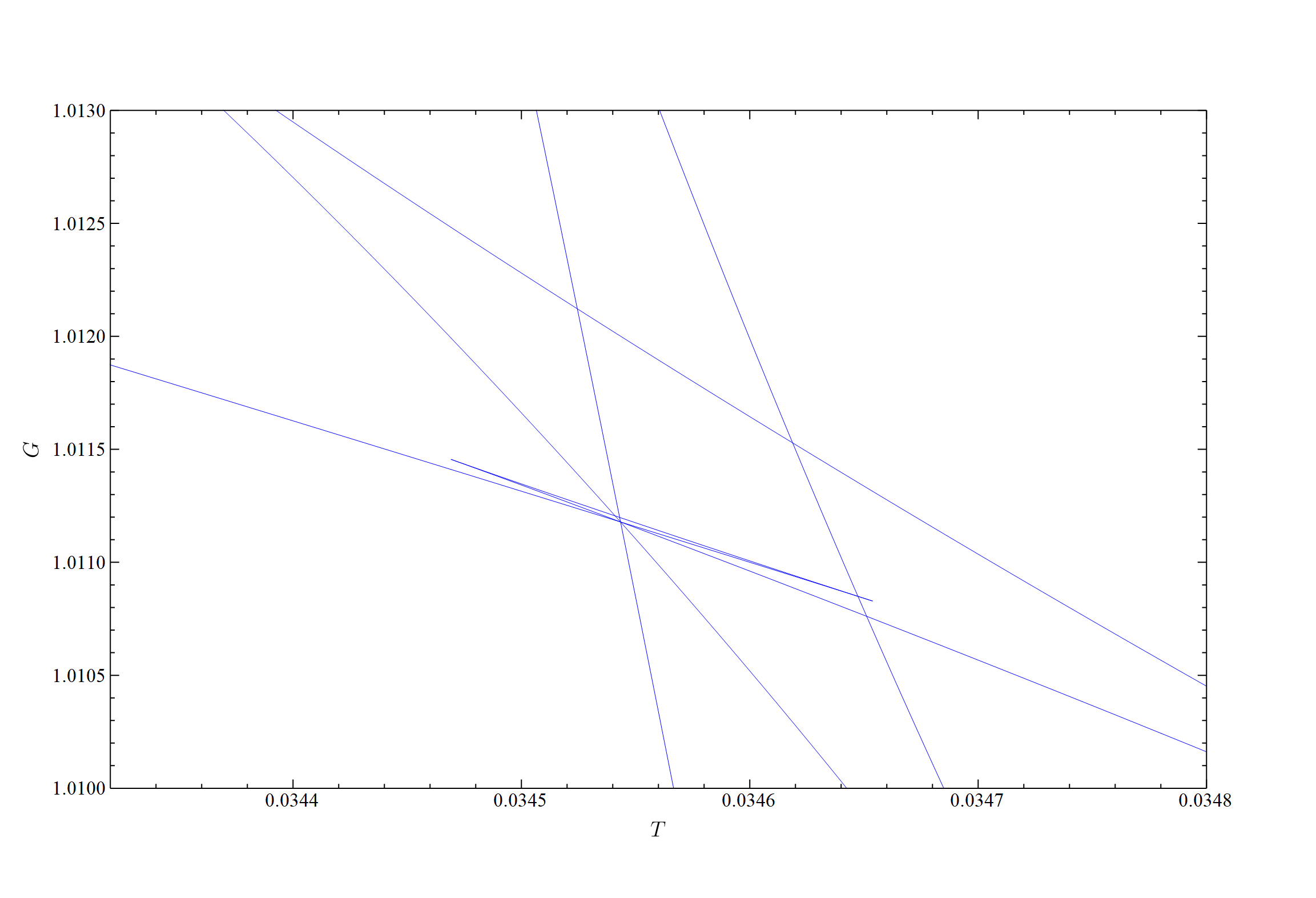}
\caption{\label{fig:PMW2}   The figure shows the Gibbs free energy corresponds to the equation of states \eqref{tempEPM} with $Q=1.2431, P^*=0.0020692, b_5=-1.1285,b_9=1.6683,b_{13}=-1.3512,b_{17}=0.42992$ 
which intersects with $T^*=0.034544$ at $r_+^{(1)}=1,r_+^{(2)}=1.0424,r_+^{(3)}=1.1,r_+^{(4)}=1.4243,r_+^{(5)}=1.9,r_+^{(6)}=2.93,r_+^{(7)}=4$.  
}

\end{sidewaysfigure}

\clearpage

\appendix   
\section{Physical constraints}\label{sec:constraints}

The equation of motion for $f(r)$ with non-zero black hole mass admits asymptotically AdS solutions of the form 
\begin{equation}\label{farsol-app}
    f(r)=\kappa+f_{\infty}\frac{r^2}{\ell^2}+\frac{m}{h'(f_{\infty})r^{d-3}}+\cdots,
\end{equation}
with omitted terms decaying faster than $1/r^{d-3}$. The parameter $m$ is related to the ADM mass $M$ via 
\begin{equation*}
M=\frac{(d-2)\Omega_{d-2,\kappa}}{16\pi}m.
\end{equation*}
The constant $f_{\infty}$ satisfies the polynomial equation
\begin{equation}\label{embeddingfunction}
    h(f_{\infty})\equiv 1-f_{\infty}+\sum_{n\geq 2}^{n_{\mathrm{max}}}\sum_{k=1}^{k_p}\frac{(-1)^{n}2\alpha_{n,k}}{\ell^{2n-2}} \frac{d-2n}{d-2}f^n_{\infty}=0,
\end{equation}
whose first derivative  with respect to $f_{\infty}$ is
\begin{equation}
    h'(f_{\infty})=-1+\sum_{n\geq 2}^{n_{\mathrm{max}}}\sum_{k=1}^{k_p}\frac{(-1)^n2\alpha_{n,k}}{\ell^{2n-2}} \frac{d-2n}{d-2}nf^{n-1}_{\infty}.
\end{equation}
These conditions ensure that the graviton is not a ghost in a constant curvature background. In some theories $h'(f_{\infty})$ vanishes identically \cite{Arenas-Henriquez:2017xnr,Arenas-Henriquez:2019rph}, in which cases the asymptotic expansion \eqref{farsol} is no longer valid. We shall not consider this situation here, leaving it   for future investigation.

For the QT class, a positive value of $f_{\infty}$ that solves \eqref{embeddingfunction} is sufficient to impose asymptotically AdS  boundary conditions, but  is insufficient for a theory with genuine GQT densities. This is because of the non-algebraic nature of equation \eqref{eom}.  Since this is a differential equation, the expansion \eqref{farsol} only represents one particular solution for $f(r)$.  To gain a complete version of the asymptotic solution, we must find the homogenous part $f_h$, which solves the differential equation
\begin{equation}\label{homoeqn}
    f''_h-\frac{4}{r}f'_h-\gamma^2r^{d-3}f_h=0.
\end{equation}
Although the structure of \eqref{homoeqn} is general for any genuine GQT theories, the expression
for the coefficient $\gamma^2$ here depends on the choice of higher-curvature couplings. For example, the expression 
\begin{equation}\label{gamma21-app}
    \gamma^2=\frac{4[ h'(f_{\infty})]^2  f_{\infty}^2}{3 (d-1) m [ h(f_{\infty})+f_{\infty}-1]\ell^2} 
\end{equation}
holds for  theories that have $\{\alpha_{n,k=2}\}$ as the only non-zero couplings, but   becomes invalid if any coupling with $k\geq3$ is turned on. 

Hence for GQT theories, in addition to the sign of $f_{\infty}$, we require that $f_h$ have only decaying modes faster than $1/r^{d-3}$ to preserve the correct asymptotic condition at the boundary. Solving \eqref{homoeqn} at far region yields
\begin{equation*}
    f_h=
    \begin{cases}
        r^{5/2} \left[A I_{\frac{5}{(d - 1)}}\left( 2\gamma \frac{r^{(d-1)/2 }}{d - 1} \right)
+ B K_ {\frac{5}{(d - 1)}}\left( 2\gamma \frac{r^{(d-1)/2 }}{d - 1} \right)
\right]& \text{if $\gamma^2>0$}\\
\\
        r^{5/2} \left[C J_{\frac{5}{(d - 1)}}\left( 2\gamma \frac{r^{(d-1)/2 }}{d - 1} \right)
+ D Y_ {\frac{5}{(d - 1)}}\left( 2\gamma \frac{r^{(d-1)/2 }}{d - 1} \right)
\right]& \text{if $\gamma^2<0$},
    \end{cases}
\end{equation*}
where $A$, $B$, $C$ and $D$ are constants fixed by boundary conditions and $I_k(x)$, $K_k(x)$, $ J_k(x)$ and $Y_k(x)$ are  Bessel functions of order $k$. In the large-$r$ limit they become
\begin{equation*}
      f_h^{(+)}\sim Ar^{\frac{11-d}{4}}\exp\bigg(\frac{2|\gamma| r^{\frac{d-1}{2}}}{d-1}\bigg)+Br^{\frac{11-d}{4}}\exp\bigg(-\frac{2|\gamma| r^{\frac{d-1}{2}}}{d-1}\bigg),
\end{equation*}
\begin{equation*}
     f_h^{(-)}\sim C r^{\frac{11-d}{4}} \cos{\bigg(\frac{2|\gamma|r^{\frac{d-1}{2}}}{d-1}+\frac{(3d-13)\pi}{4(d-1)}\bigg)}+D r^{\frac{11-d}{4}} \sin{\bigg(\frac{2|\gamma|r^{\frac{d-1}{2}}}{d-1}+\frac{(3d-13)\pi}{4(d-1)}\bigg)},
\end{equation*}
where  the superscripts $(\pm)$ indicate the sign of $\gamma^2$. In order that the homogenous part is subdominant at large $r$,  we require $\gamma^2 > 0$ and $A=0$. 

Apart from the correct asymptote, physical theories should only propagate one type of massless spin-2 graviton on constant curvature backgrounds. This implies the effective Newtonian constant must have the same sign as the one in general relativity, which means the third term in \eqref{farsol} should become negative for positive mass, that is  $h'(f_{\infty})<0$ \cite{hennigar2017generalized}.

Summarizing, we only consider black holes with 
$f_{\infty}>0$, $  h'(f_{\infty})<0 $ and $ \gamma^2>0$ as satisfying the requisite physical criteria.

\section{Einstein-power-Maxwell AdS black holes}

Here we compare our method with methods previously employed in understanding multicritical behaviour.
 For this illustration,  we consider a quadruple point in the Einstein-power-Maxwell theory 
 comparing prevous methods used to find it
 \cite{tavakoli2022multi} with our proposed approach.

Power-Maxwell theory is 
a general form of 
non-linear electrodynamics minimally coupled to $D=4$ Einstein gravity \cite{Gao:2021kvr}. The action   is  
\begin{equation}\label{action}
	S=\int d^4x \sqrt{-g}\left(R-2 \Lambda -\sum_{i=1}^{N} \alpha _i (F^2)^i \right),
\end{equation}
with $F^2 \equiv F_{\mu \nu }F^{\mu \nu}$ and $F_{\mu \nu}\equiv \nabla_{\mu} A_\nu - \nabla _\nu A_\nu$, where
  $R$ is the Ricci scalar. The $ \alpha _i$ are dimensional coupling constants ( $[\alpha_i]=L^{2(i-1)}$)  and $A_\mu$ is the U(1) Maxwell field. We recover Einstein-Maxwell theory when $\alpha_1=1$ and $\alpha_i (i>1) = 0$.  
 
For the ansatz \eqref{sssansatz} it is straightforward to find a solution of the form
$f(r) = 1+\sum_{i=1}^{K} c_i r^{-i}+\frac{r^2}{l^2}$
\cite{tavakoli2022multi}. The corresponding thermodynamic variables are
\begin{equation}\label{tempEPM}
\begin{aligned}
T=\frac{1}{4\pi r_+}&\bigg[1+8\pi P r_+^2-\frac{Q^2}{r_+^2}-\frac{5b_5 Q}{2r_+^6}-\frac{3b_9 Q}{r_+^{10}}\\&-\frac{13b_{13}Q}{4r_+^{14}}-\frac{17b_{17}Q}{5r_+^{18}}-\frac{7b_{21}Q}{5r_+^{22}}+\mathcal{O}\bigg(\frac{1}{r_+^{26}}\bigg)\bigg],
\end{aligned}
\end{equation}
\begin{equation}
    S=\pi r_+^2, \quad V=\frac{4}{3}\pi r_+^3, \quad P=\frac{3}{8\pi l^2},
\end{equation}
where $Q$ is the charge parameter,  and each of the other parameters represent the same physical quantities as before. The field equations imply
\begin{equation}
   c_1= - 2M
    \qquad
  c_i=\frac{4Q}{i+2}b_{i-1},
    \qquad \text{for}  \qquad
    i>1
  \end{equation}
where
\begin{align}
	&b_1=Q \qquad 
	b_5=\frac{4}{5}Q^3\alpha_2
	\qquad
	b_9=\frac{4}{3}Q^5(4\alpha_2^2-\alpha_3) 
	\\
	&b_{13} = \frac{32}{13}Q^7(24 \alpha_2^3-12 \alpha_3\alpha_2+\alpha_4)
	\\
	&b_{17}=\frac{80}{17}Q^9(1
	76\alpha_2^4-132 \alpha_2^2\alpha_3+16\alpha_4\alpha_2+9\alpha_3^2-\alpha_5)
	\\
	&b_{21}=\frac{64}{7}Q^{11}(1456\alpha_2^5+234\alpha_3^2\alpha_2+208\alpha_4\alpha_2^2-24\alpha_4\alpha_3
	\nonumber\\
	&\qquad -1456\alpha_2^3\alpha_3-20\alpha_5\alpha_2+\alpha_6
	)
\end{align}
and we can set $\alpha_1 =1$ without loss of generality.  

The Gibbs free energy is again $G=M-TS$ but with the black hole mass now modified to be
\begin{equation}
\begin{aligned}
    M=& \frac{4}{3} \pi  P r_+^3+\frac{r_+}{2}+\frac{Q^2}{2 r_+}+\frac{b_5 Q}{4 r_+^5}+\frac{b_9
   Q}{6 r_+^9}\\&+\frac{b_{13} Q}{8 r_+^{13}}+\frac{b_{17} Q}{10 r_+^{17}}+\frac{b_{21}Q}{12r_+^{21}}+\mathcal{O}\bigg(\frac{1}{r_+^{25}}\bigg).
\end{aligned}
\end{equation}

The approach\footnote{There is another method mentioned in the paper that uses extrema. However, they are qualitatively the same since neither of them avoid fine-tuning.} stated in \cite{tavakoli2022multi} requires presetting $2N-1$ intersections $r_+^{(i)}$ to construct an $N$-tuple point by solving $2N-1$ equations 
\begin{equation}
    T(r_+^{(1)},P^*,\{b_n\})=\cdots=T(r_+^{(2N-1)},P^*,\{b_n\})=T^*
\end{equation}
for $T^*$, $P^*$, $Q$, $b_n$. This yields a numerical expression for the temperature \eqref{tempEPM} each time we change the values of intersections. Hence an $N$-tuple point  (likely) emerges eventually by adjusting the $r_+^{(i)}$ such that $N-1$ swallowtails of the free enthapy appears at a single point. The feasibility of the method is lucid by applying a similar analysis in the section \ref{sec:construction} to \eqref{tempEPM}. As a consequence of the rule of signs, $2N-4$ is the minimum number of couplings mandatory to construct an $N$-tuple point. 

For an illustration, we apply this approach to a quadruple point, and so four non-zero $b_n$'s are required. For simplicity, we choose them to be $\{b_5,b_9,b_{13},b_{17}\}$. In order to work out a desired set of $b_n$'s, we pick intersections $\{1,1.04,1.1,1.4,1.9,2.9,4\}$ (set A) to start with. However, this choice of $r_+^{(i)}$ is not optimal in the sense that only two swallowtails appear in the Gibbs free energy, as shown in  figure \ref{fig:PMW0}, which is far from   a quadruple point merger. After a few iterations, we arrive at a better situation displayed by the figure \ref{fig:PMW1}, where all three swallowtails appear. These do  not get intersect a single point until all radii are tuned to $\{1,1.0424,1.1,1.4243,1.9,2.93,4\}$ (set B) as illustrated in   figure \ref{fig:PMW2}. Radii with  higher precision can be obtained if one iterates the approach for many times, but a considerably longer time is inevitably needed.

We now consider how things change using our method. We first define the function $K$ as
\begin{equation}
    K(r_+,r_0)=\int_{r_0}^{r_+}4\pi\bigg[P^*-P(\Tilde{r}_+,T^*,\{b_n\})\bigg] \Tilde{r}_+^2 \mathrm{d} \Tilde{r}
\end{equation}
where \eqref{tempEPM} and $V=4\pi r_+^3/3$ are applied to \eqref{myK}. Next pick $\{1,1.1,1.9,4\}$ out of the set A to be four different values of $r_+$ satisfying \eqref{Krule}. After one implementation of the procedure discussed in section \ref{sec:construction}, we find 
\begin{equation}
\begin{aligned}
    &\qquad\qquad\quad r_+^{(1)}=1, \quad r_+^{(2)}=1.042427218, \quad r_+^{(3)}=1.1, \\& r_+^{(4)}=1.424291548, \quad r_+^{(5)}=1.9, \quad r_+^{(6)}=2.930025838, \quad r_+^{(7)}=4,
\end{aligned}
\end{equation}
which reproduces the set B with considerably higher precision and notably less effort. From this perspective, our approach can be interpreted as a fine-tuning process that automatically identifies all possible proper positions of the $N-1$ remaining radii in between the preset $N$ radii such that an $N$-tuple point emerges. All those radii collectively constitute a workable set of $r_+^{(i)}$ which constructs an $N$-tuple point in the preceding method.


\bibliographystyle{JHEP}
\bibliography{ref.bib}

\providecommand{\href}[2]{#2}\begingroup\raggedright\begin{thebibliography}{10}

\bibitem{hawking1974black}
S.~W. Hawking, {\it {Black hole explosions}},  {\em Nature} {\bf 248} (1974)
  30--31.

\bibitem{bardeen1973four}
J.~M. Bardeen, B.~Carter, and S.~W. Hawking, {\it {The Four laws of black hole
  mechanics}},  {\em Commun. Math. Phys.} {\bf 31} (1973) 161--170.

\bibitem{creighton1995quasilocal}
J.~D.~E. Creighton and R.~B. Mann, {\it {Quasilocal thermodynamics of dilaton
  gravity coupled to gauge fields}},  {\em Phys. Rev. D} {\bf 52} (1995)
  4569--4587, [\href{http://arxiv.org/abs/gr-qc/9505007}{{\tt gr-qc/9505007}}].

\bibitem{henneaux1985asymptotically}
M.~Henneaux and C.~Teitelboim, {\it {Asymptotically anti-De Sitter Spaces}},
  {\em Commun. Math. Phys.} {\bf 98} (1985) 391--424.

\bibitem{kastor2010smarr}
D.~Kastor, S.~Ray, and J.~Traschen, {\it {Smarr Formula and an Extended First
  Law for Lovelock Gravity}},  {\em Class. Quant. Grav.} {\bf 27} (2010)
  235014, [\href{http://arxiv.org/abs/1005.5053}{{\tt arXiv:1005.5053}}].

\bibitem{kastor2011mass}
D.~Kastor, S.~Ray, and J.~Traschen, {\it {Mass and Free Energy of Lovelock
  Black Holes}},  {\em Class. Quant. Grav.} {\bf 28} (2011) 195022,
  [\href{http://arxiv.org/abs/1106.2764}{{\tt arXiv:1106.2764}}].

\bibitem{kubizvnak2015black}
D.~Kubiznak and R.~B. Mann, {\it {Black hole chemistry}},  {\em Can. J. Phys.}
  {\bf 93} (2015), no.~9 999--1002, [\href{http://arxiv.org/abs/1404.2126}{{\tt
  arXiv:1404.2126}}].

\bibitem{kubizvnak2017black}
D.~Kubiznak, R.~B. Mann, and M.~Teo, {\it {Black hole chemistry: thermodynamics
  with Lambda}},  {\em Class. Quant. Grav.} {\bf 34} (2017), no.~6 063001,
  [\href{http://arxiv.org/abs/1608.06147}{{\tt arXiv:1608.06147}}].

\bibitem{kubizvnak2012p}
D.~Kubiznak and R.~B. Mann, {\it {P-V criticality of charged AdS black holes}},
   {\em JHEP} {\bf 07} (2012) 033, [\href{http://arxiv.org/abs/1205.0559}{{\tt
  arXiv:1205.0559}}].

\bibitem{altamirano2013reentrant}
N.~Altamirano, D.~Kubiznak, and R.~B. Mann, {\it {Reentrant phase transitions
  in rotating anti\textendash{}de Sitter black holes}},  {\em Phys. Rev. D}
  {\bf 88} (2013), no.~10 101502, [\href{http://arxiv.org/abs/1306.5756}{{\tt
  arXiv:1306.5756}}].

\bibitem{frassino2014multiple}
A.~M. Frassino, D.~Kubiznak, R.~B. Mann, and F.~Simovic, {\it {Multiple
  Reentrant Phase Transitions and Triple Points in Lovelock Thermodynamics}},
  {\em JHEP} {\bf 09} (2014) 080, [\href{http://arxiv.org/abs/1406.7015}{{\tt
  arXiv:1406.7015}}].

\bibitem{dykaar2017hairy}
H.~Dykaar, R.~A. Hennigar, and R.~B. Mann, {\it {Hairy black holes in cubic
  quasi-topological gravity}},  {\em JHEP} {\bf 05} (2017) 045,
  [\href{http://arxiv.org/abs/1703.01633}{{\tt arXiv:1703.01633}}].

\bibitem{hennigar2017superfluid}
R.~A. Hennigar, R.~B. Mann, and E.~Tjoa, {\it {Superfluid Black Holes}},  {\em
  Phys. Rev. Lett.} {\bf 118} (2017), no.~2 021301,
  [\href{http://arxiv.org/abs/1609.02564}{{\tt arXiv:1609.02564}}].

\bibitem{hennigar2017thermodynamics}
R.~A. Hennigar, E.~Tjoa, and R.~B. Mann, {\it {Thermodynamics of hairy black
  holes in Lovelock gravity}},  {\em JHEP} {\bf 02} (2017) 070,
  [\href{http://arxiv.org/abs/1612.06852}{{\tt arXiv:1612.06852}}].

\bibitem{altamirano2014kerr}
N.~Altamirano, D.~Kubiz\v{n}\'ak, R.~B. Mann, and Z.~Sherkatghanad, {\it
  {Kerr-AdS analogue of triple point and solid/liquid/gas phase transition}},
  {\em Class. Quant. Grav.} {\bf 31} (2014) 042001,
  [\href{http://arxiv.org/abs/1308.2672}{{\tt arXiv:1308.2672}}].

\bibitem{hull2021thermodynamics}
B.~R. Hull and R.~B. Mann, {\it {Thermodynamics of exotic black holes in
  Lovelock gravity}},  {\em Phys. Rev. D} {\bf 104} (2021), no.~8 084032,
  [\href{http://arxiv.org/abs/2102.05282}{{\tt arXiv:2102.05282}}].

\bibitem{hull2022exotic}
B.~R. Hull and F.~Simovic, {\it {Exotic Black Hole Thermodynamics in
  Third-Order Lovelock Gravity}},  \href{http://arxiv.org/abs/2208.05500}{{\tt
  arXiv:2208.05500}}.

\bibitem{tavakoli2022multi}
M.~Tavakoli, J.~Wu, and R.~B. Mann, {\it {Multi-critical points in black hole
  phase transitions}},  {\em JHEP} {\bf 12} (2022) 117,
  [\href{http://arxiv.org/abs/2207.03505}{{\tt arXiv:2207.03505}}].

\bibitem{wu2022multicritical}
J.~Wu and R.~B. Mann, {\it {Multicritical phase transitions in multiply
  rotating black holes}},  {\em Class. Quant. Grav.} {\bf 40} (2023), no.~6
  06LT01, [\href{http://arxiv.org/abs/2208.00012}{{\tt arXiv:2208.00012}}].

\bibitem{wu2022multicritical2}
J.~Wu and R.~B. Mann, {\it {Multicritical phase transitions in Lovelock AdS
  black holes}},  {\em Phys. Rev. D} {\bf 107} (2023), no.~8 084035,
  [\href{http://arxiv.org/abs/2212.08087}{{\tt arXiv:2212.08087}}].

\bibitem{wu2022thermodynamically}
J.~Wu and R.~B. Mann, {\it {Thermodynamically Stable Phases of Asymptotically
  Flat Lovelock Black Holes}},  \href{http://arxiv.org/abs/2212.08673}{{\tt
  arXiv:2212.08673}}.

\bibitem{clerk1875dynamical}
J.~Clerk-Maxwell, {\it On the dynamical evidence of the molecular constitution
  of bodies},  {\em Nature} {\bf 11} (1875), no.~279 357--359.

\bibitem{stelle1977renormalization}
K.~S. Stelle, {\it {Renormalization of Higher Derivative Quantum Gravity}},
  {\em Phys. Rev. D} {\bf 16} (1977) 953--969.

\bibitem{woodard2015ostrogradsky}
R.~P. Woodard, {\it {Ostrogradsky's theorem on Hamiltonian instability}},  {\em
  Scholarpedia} {\bf 10} (2015), no.~8 32243,
  [\href{http://arxiv.org/abs/1506.02210}{{\tt arXiv:1506.02210}}].

\bibitem{OstrogradskyMmoiresSL}
M.~Ostrogradsky, {\it M{\'e}moires sur les {\'e}quations diff{\'e}rentielles,
  relatives au probl{\`e}me des isop{\'e}rim{\`e}tres}, .

\bibitem{lovelock1970divergence}
D.~Lovelock, {\it Divergence-free tensorial concomitants},  {\em Aequationes
  mathematicae} {\bf 4} (1970), no.~1 127--138.

\bibitem{lovelock1971einstein}
D.~Lovelock, {\it {The Einstein tensor and its generalizations}},  {\em J.
  Math. Phys.} {\bf 12} (1971) 498--501.

\bibitem{sotiriou2010f}
T.~P. Sotiriou and V.~Faraoni, {\it {f(R) Theories Of Gravity}},  {\em Rev.
  Mod. Phys.} {\bf 82} (2010) 451--497,
  [\href{http://arxiv.org/abs/0805.1726}{{\tt arXiv:0805.1726}}].

\bibitem{woodard2007avoiding}
R.~P. Woodard, {\it {Avoiding dark energy with 1/r modifications of gravity}},
  {\em Lect. Notes Phys.} {\bf 720} (2007) 403--433,
  [\href{http://arxiv.org/abs/astro-ph/0601672}{{\tt astro-ph/0601672}}].

\bibitem{bueno2019generalized}
P.~Bueno, P.~A. Cano, and R.~A. Hennigar, {\it {(Generalized) quasi-topological
  gravities at all orders}},  {\em Class. Quant. Grav.} {\bf 37} (2020), no.~1
  015002, [\href{http://arxiv.org/abs/1909.07983}{{\tt arXiv:1909.07983}}].

\bibitem{Bueno:2022res}
P.~Bueno, P.~A. Cano, R.~A. Hennigar, M.~Lu, and J.~Moreno, {\it {Generalized
  quasi-topological gravities: the whole shebang}},  {\em Class. Quant. Grav.}
  {\bf 40} (2023), no.~1 015004, [\href{http://arxiv.org/abs/2203.05589}{{\tt
  arXiv:2203.05589}}].

\bibitem{Moreno:2023rfl}
J.~Moreno and A.~J. Murcia, {\it {On the classification of Generalized
  Quasitopological Gravities}},  \href{http://arxiv.org/abs/2304.08510}{{\tt
  arXiv:2304.08510}}.

\bibitem{myers2010holographic}
R.~C. Myers, M.~F. Paulos, and A.~Sinha, {\it {Holographic studies of
  quasi-topological gravity}},  {\em JHEP} {\bf 08} (2010) 035,
  [\href{http://arxiv.org/abs/1004.2055}{{\tt arXiv:1004.2055}}].

\bibitem{myers2010black}
R.~C. Myers and B.~Robinson, {\it {Black Holes in Quasi-topological Gravity}},
  {\em JHEP} {\bf 08} (2010) 067, [\href{http://arxiv.org/abs/1003.5357}{{\tt
  arXiv:1003.5357}}].

\bibitem{oliva2010new}
J.~Oliva and S.~Ray, {\it {A new cubic theory of gravity in five dimensions:
  Black hole, Birkhoff's theorem and C-function}},  {\em Class. Quant. Grav.}
  {\bf 27} (2010) 225002, [\href{http://arxiv.org/abs/1003.4773}{{\tt
  arXiv:1003.4773}}].

\bibitem{cisterna2017quintic}
A.~Cisterna, L.~Guajardo, M.~Hassaine, and J.~Oliva, {\it {Quintic
  quasi-topological gravity}},  {\em JHEP} {\bf 04} (2017) 066,
  [\href{http://arxiv.org/abs/1702.04676}{{\tt arXiv:1702.04676}}].

\bibitem{oliva2011birkhoff}
J.~Oliva and S.~Ray, {\it {Birkhoff's Theorem in Higher Derivative Theories of
  Gravity}},  {\em Class. Quant. Grav.} {\bf 28} (2011) 175007,
  [\href{http://arxiv.org/abs/1104.1205}{{\tt arXiv:1104.1205}}].

\bibitem{oliva2012birkhoff}
J.~Oliva and S.~Ray, {\it {Birkhoff's Theorem in Higher Derivative Theories of
  Gravity II}},  {\em Phys. Rev. D} {\bf 86} (2012) 084014,
  [\href{http://arxiv.org/abs/1201.5601}{{\tt arXiv:1201.5601}}].

\bibitem{bueno2017black}
P.~Bueno and P.~A. Cano, {\it {On black holes in higher-derivative gravities}},
   {\em Class. Quant. Grav.} {\bf 34} (2017), no.~17 175008,
  [\href{http://arxiv.org/abs/1703.04625}{{\tt arXiv:1703.04625}}].

\bibitem{bueno2017universal}
P.~Bueno and P.~A. Cano, {\it {Universal black hole stability in four
  dimensions}},  {\em Phys. Rev. D} {\bf 96} (2017), no.~2 024034,
  [\href{http://arxiv.org/abs/1704.02967}{{\tt arXiv:1704.02967}}].

\bibitem{hennigar2017criticality}
R.~A. Hennigar, {\it {Criticality for charged black branes}},  {\em JHEP} {\bf
  09} (2017) 082, [\href{http://arxiv.org/abs/1705.07094}{{\tt
  arXiv:1705.07094}}].

\bibitem{hennigar2017generalized}
R.~A. Hennigar, D.~Kubiz\v{n}\'ak, and R.~B. Mann, {\it {Generalized
  quasitopological gravity}},  {\em Phys. Rev. D} {\bf 95} (2017), no.~10
  104042, [\href{http://arxiv.org/abs/1703.01631}{{\tt arXiv:1703.01631}}].

\bibitem{mir2019generalized}
M.~Mir and R.~B. Mann, {\it {On generalized quasi-topological cubic-quartic
  gravity: thermodynamics and holography}},  {\em JHEP} {\bf 07} (2019) 012,
  [\href{http://arxiv.org/abs/1902.10906}{{\tt arXiv:1902.10906}}].

\bibitem{mir2019black}
M.~Mir, R.~A. Hennigar, J.~Ahmed, and R.~B. Mann, {\it {Black hole chemistry
  and holography in generalized quasi-topological gravity}},  {\em JHEP} {\bf
  08} (2019) 068, [\href{http://arxiv.org/abs/1902.02005}{{\tt
  arXiv:1902.02005}}].

\bibitem{Deser:2002jk}
S.~Deser and B.~Tekin, {\it {Energy in generic higher curvature gravity
  theories}},  {\em Phys. Rev. D} {\bf 67} (2003) 084009,
  [\href{http://arxiv.org/abs/hep-th/0212292}{{\tt hep-th/0212292}}].

\bibitem{Arnowitt:1960zzc}
R.~Arnowitt, S.~Deser, and C.~W. Misner, {\it {Energy and the Criteria for
  Radiation in General Relativity}},  {\em Phys. Rev.} {\bf 118} (1960)
  1100--1104.

\bibitem{Arnowitt:1961zz}
R.~L. Arnowitt, S.~Deser, and C.~W. Misner, {\it {Coordinate invariance and
  energy expressions in general relativity}},  {\em Phys. Rev.} {\bf 122}
  (1961) 997.

\bibitem{Arnowitt:1960es}
R.~L. Arnowitt, S.~Deser, and C.~W. Misner, {\it {Canonical variables for
  general relativity}},  {\em Phys. Rev.} {\bf 117} (1960) 1595--1602.

\bibitem{Wald:1993nt}
R.~M. Wald, {\it {Black hole entropy is the Noether charge}},  {\em Phys. Rev.
  D} {\bf 48} (1993), no.~8 R3427--R3431,
  [\href{http://arxiv.org/abs/gr-qc/9307038}{{\tt gr-qc/9307038}}].

\bibitem{peters2020defying}
V.~Peters, M.~Vis, {\'A}.~G. Garc{\'\i}a, H.~H. Wensink, and R.~Tuinier, {\it
  Defying the gibbs phase rule: Evidence for an entropy-driven quintuple point
  in colloid-polymer mixtures},  {\em Physical Review Letters} {\bf 125}
  (2020), no.~12 127803.

\bibitem{garcia2018depletion}
{\'A}.~G. Garc{\'\i}a, R.~Tuinier, J.~V. Maring, J.~Opdam, H.~H. Wensink, and
  H.~N. Lekkerkerker, {\it Depletion-driven four-phase coexistences in discotic
  systems},  {\em Molecular Physics} {\bf 116} (2018), no.~21-22 2757--2772.

\bibitem{Wei:2019uqg}
S.-W. Wei, Y.-X. Liu, and R.~B. Mann, {\it {Repulsive Interactions and
  Universal Properties of Charged Anti\textendash{}de Sitter Black Hole
  Microstructures}},  {\em Phys. Rev. Lett.} {\bf 123} (2019), no.~7 071103,
  [\href{http://arxiv.org/abs/1906.10840}{{\tt arXiv:1906.10840}}].

\bibitem{Wei:2019yvs}
S.-W. Wei, Y.-X. Liu, and R.~B. Mann, {\it {Ruppeiner Geometry, Phase
  Transitions, and the Microstructure of Charged AdS Black Holes}},  {\em Phys.
  Rev. D} {\bf 100} (2019), no.~12 124033,
  [\href{http://arxiv.org/abs/1909.03887}{{\tt arXiv:1909.03887}}].

\bibitem{Hale:2023qjx}
T.~Hale, D.~Kubiznak, O.~Svitek, and T.~Tahamtan, {\it {MadMax electrodynamics:
  Solutions and basic properties}},
  \href{http://arxiv.org/abs/2303.16928}{{\tt arXiv:2303.16928}}.

\bibitem{Caceres:2015vsa}
E.~Caceres, P.~H. Nguyen, and J.~F. Pedraza, {\it {Holographic entanglement
  entropy and the extended phase structure of STU black holes}},  {\em JHEP}
  {\bf 09} (2015) 184, [\href{http://arxiv.org/abs/1507.06069}{{\tt
  arXiv:1507.06069}}].

\bibitem{Arenas-Henriquez:2017xnr}
G.~Arenas-Henriquez, O.~Miskovic, and R.~Olea, {\it {Vacuum Degeneracy and
  Conformal Mass in Lovelock AdS Gravity}},  {\em JHEP} {\bf 11} (2017) 128,
  [\href{http://arxiv.org/abs/1710.08512}{{\tt arXiv:1710.08512}}].

\bibitem{Arenas-Henriquez:2019rph}
G.~Arenas-Henriquez, R.~B. Mann, O.~Miskovic, and R.~Olea, {\it {Mass in
  Lovelock Unique Vacuum gravity theories}},  {\em Phys. Rev. D} {\bf 100}
  (2019), no.~6 064038, [\href{http://arxiv.org/abs/1905.10840}{{\tt
  arXiv:1905.10840}}].

\bibitem{Gao:2021kvr}
C.~Gao, {\it {Black holes with many horizons in the theories of nonlinear
  electrodynamics}},  {\em Phys. Rev. D} {\bf 104} (2021), no.~6 064038,
  [\href{http://arxiv.org/abs/2106.13486}{{\tt arXiv:2106.13486}}].

\end{thebibliography}\endgroup

\end{document}